\documentclass[twocolumn,tighten,times,twocolappendix,numberedappendix]{aastex631}

\usepackage{amsmath}
\usepackage{float}
\usepackage{color}
\usepackage{url}
\usepackage{etoolbox}
\usepackage{graphicx}

\usepackage[flushleft]{threeparttable}

\hypersetup{breaklinks=true,colorlinks=true,linkcolor=blue,citecolor=blue,filecolor=blue,urlcolor=blue}
\makeatletter
\patchcmd{\NAT@citex}
  {\@citea\NAT@hyper@{%
     \NAT@nmfmt{\NAT@nm}%
     \hyper@natlinkbreak{\NAT@aysep\NAT@spacechar}{\@citeb\@extra@b@citeb}%
     \NAT@date}}
  {\@citea\NAT@nmfmt{\NAT@nm}%
   \NAT@aysep\NAT@spacechar\NAT@hyper@{\NAT@date}}{}{}

\patchcmd{\NAT@citex}
  {\@citea\NAT@hyper@{%
     \NAT@nmfmt{\NAT@nm}%
     \hyper@natlinkbreak{\NAT@spacechar\NAT@@open\if*#1*\else#1\NAT@spacechar\fi}%
       {\@citeb\@extra@b@citeb}%
     \NAT@date}}
  {\@citea\NAT@nmfmt{\NAT@nm}%
   \NAT@spacechar\NAT@@open\if*#1*\else#1\NAT@spacechar\fi\NAT@hyper@{\NAT@date}}
  {}{}

\usepackage{array}
\newcolumntype{C}[1]{>{\centering\let\newline\\\arraybackslash\hspace{0pt}}m{#1}}

\usepackage{xspace}

\newcommand{\orcidicon}{\includegraphics[width=0.26cm]{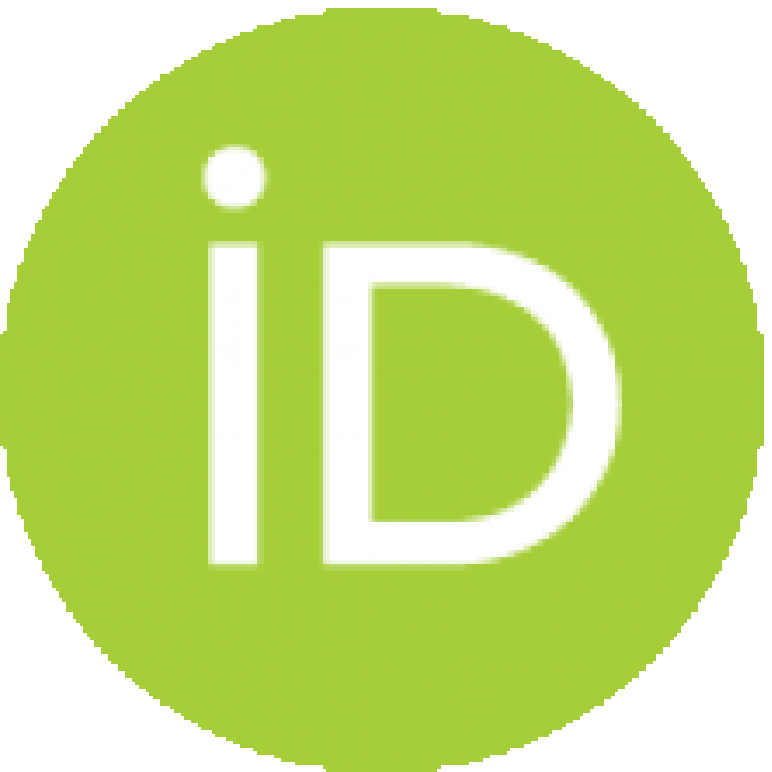}}

\newcommand{\orcidauthor}[1]{\href{https://orcid.org/#1}{\orcidicon}}

\def\apj{ApJ}
\def\apjl{ApJ}
\def\apjs{ApJS}
\def\ao{Appl.~Opt.}

\def\aap{A\&A}
\def\aapr{A\&A~Rev.}
\def\aaps{A\&AS}

\def\mnras{MNRAS}

\def\pasp{PASP}

\def\nat{Nature}

\def\rmxaa{RMxAA}

\newcommand{\maihem}{\textsc{maihem}\xspace}
\newcommand{\flash}{\textsc{flash}\xspace}
\newcommand{\cloudy}{\textsc{cloudy}\xspace}

\newcommand{\ionic}[2]{#1$\,${\scshape{#2}}\xspace}
\newcommand{\ionf}[2]{#1$\,${\scshape{#2}}}

\def\arcdeg{\hbox{$^\circ$}}

\newcommand{\ovi}{O\,{\sc vi}\xspace}

\newcommand{\civ}{C\,{\sc iv}\xspace}

\newcommand{\hii}{H\,{\sc ii}\xspace}

\newcommand{\heii}{He\,{\sc ii}\xspace}

\makeatletter
\patchcmd{\frontmatter@RRAP@format}{(}{}{}{}
\patchcmd{\frontmatter@RRAP@format}{)}{}{}{}
\renewcommand\Dated@name{}
\makeatother

\renewcommand{\vec}[1]{\mathbf{#1}}

\shorttitle{Catastrophic Cooling in Superwinds. III. Non-equilibrium Photoionization}
\shortauthors{Danehkar et al.}
\graphicspath{{./}{figures/}}

\begin{document}

\title{Catastrophic Cooling in Superwinds. III. Non-equilibrium Photoionization}

\correspondingauthor{A.~Danehkar}
\email{danehkar@eurekasci.com}

\author[0000-0003-4552-5997]{A.~Danehkar}
\affiliation{Eureka Scientific, Inc., 2452 Delmer Street Suite 100, Oakland, CA 94602-3017, USA}

\author[0000-0002-5808-1320]{M. S. Oey}
\affiliation{Department of Astronomy, University of Michigan, 1085 S. University Ave, Ann Arbor, MI 48109, USA}

\author[0000-0001-9014-3125]{W. J. Gray}
\altaffiliation{~Private address.}


\date[ ]{\footnotesize\textit{Received 2022 June 1; revised 2022 July 30; accepted 2022 August 3}}

\begin{abstract}
Observations of some starburst-driven galactic superwinds suggest that strong radiative cooling could play a key role in the nature of feedback and the formation of stars and molecular gas in star-forming galaxies.  These catastrophically cooling superwinds are not adequately described by adiabatic fluid models, but they can be reproduced by incorporating non-equilibrium radiative cooling functions into the fluid model. In this work, we have employed the atomic and cooling module \maihem implemented in the framework of the \flash hydrodynamics code to simulate the formation of radiatively cooling superwinds  as well as their corresponding non-equilibrium ionization (NEI) states for various outflow parameters, gas metallicities, and ambient densities.  We employ the photoionization program \cloudy to predict radiation- and density-bounded photoionization for these radiatively cooling superwinds, and we predict UV and optical line emission. Our non-equilibrium photoionization models built with the NEI states demonstrate the enhancement of \ionic{C}{iv}, especially in metal-rich, catastrophically cooling outflows, and \ionic{O}{vi} in metal-poor ones.  
\end{abstract}

\keywords{\href{https://astrothesaurus.org/uat/572}{Galactic winds (572)};
\href{https://astrothesaurus.org/uat/1656}{Superbubbles (1656)};
\href{https://astrothesaurus.org/uat/2028}{Cooling flows (2028)};
\href{https://astrothesaurus.org/uat/1565}{Star forming regions (1565)};
\href{https://astrothesaurus.org/uat/694}{H II regions (694)};
\href{https://astrothesaurus.org/uat/1570}{Starburst galaxies (1570)};
\href{https://astrothesaurus.org/uat/459}{Emission line galaxies (459)};
\href{https://astrothesaurus.org/uat/979}{Lyman-break galaxies (979)};
\href{https://astrothesaurus.org/uat/978}{Lyman-alpha galaxies (978)}
\vspace{4pt}
\newline
\textit{Supporting material:} interactive figure, machine-readable tables}


\section{Introduction}
\label{cooling:introduction}


Observations of star-forming galaxies reveal the presence of galactic outflows, known as \textit{superwinds} \citep{Heckman1990}, with multiphase structures having regions with low temperatures ($10^{1{\rm -}3}$\,K) based on sub-millimeter and infrared observations \citep{Ott2005,Weis2005,Bolatto2013,Leroy2015}, warm ($\approx 10^{4{\rm -}4.5}$\,K) wind regions as seen in optical and near-UV measurements \citep{Walsh1989,Izotov1999,James2009,James2013}, as well as hot ($10^{6.5{\rm -}8}$\,K) bubbles in X-ray observations \citep{dellaCeca1996,Strickland1997,Martin1999,Ott2005a}. 
Historically, starburst-driven superwinds have been modeled using adiabatic assumptions \citep{Castor1975,Weaver1977,Chevalier1985,Canto2000}. 
However, adiabatic models are not consistent with suppressed
superwinds observed in several starburst galaxies such as M82
\citep{Smith2006,Westmoquette2014}, NGC\,5253 \citep{Turner2017},
NGC\,2366 \citep{Oey2017}, and extreme Green Peas
\citep[GPs;][]{Jaskot2017}. In particular, cooling superwinds can
produce virial temperatures below $10^4$\,K, and can stimulate star formation \citep{Fabian1984,Sarazin1988,Krause2016,Silich2017} and extensive molecular gas \citep[e.g.,][]{Veilleux2020}.

Semianalytic solutions for radiative superwinds, which have been
obtained with cooling functions
\citep[e.g.,][]{Silich2003,Silich2004,Tenorio-Tagle2005,Tenorio-Tagle2007},
show that temperature profiles of superwinds have significant
departures from the adiabatic solutions, and are therefore also known
as \textit{catastrophic cooling} models.
These semi-analytical results indicate that radiative cooling is
contingent on the gas metallicity, mass loading, and kinetic heating efficiency.
A semi-analytic analysis of such a model by
\citet{Wuensch2011} also demonstrates that strong radiative cooling
with low heating efficiency is more sensitive to the gas metallicity,
while the metallicity effect is not significant with large mass
loading. More recently, hydrodynamic simulations of superwinds have
also been carried out by \citet{Danehkar2021}, which confirmed the
parametric dependence of radiative cooling. Photoionization
calculations have been conducted by \citet{Danehkar2021} under the
assumption that gas is in collisional ionization equilibrium (CIE) and
photoionization equilibrium (PIE). The radiative cooling functions
have been calculated in CIE by a number of authors
\citep[][]{Cox1969,Raymond1976,Shull1982,Sutherland1993,Bryans2006}.  
The cooling functions for gas in ionization equilibrium have been employed for modeling radiatively cooling superwinds \citep[see e.g.][]{Schneider2018,Lochhaas2018,Lochhaas2020}. However, the CIE assumption is not entirely correct when the time-scale of ionization or recombination of a gas exceeds its cooling time \citep{Gnat2007,Oppenheimer2013}.


Time-dependent radiative cooling functions produce non-equilibrium ionization (NEI) states that have significant departures from CIE states \citep{Kafatos1973,Shapiro1976,Schmutzler1993,Gnat2007,Vasiliev2011,Oppenheimer2013}. 
At temperatures below $10^6$\,K, where the gas is transiting from pure CIE to PIE, NEI departures from CIE are predicted to be considerable \citep[][]{Vasiliev2011}. 
A non-equilibrium chemistry network with cooling functions was incorporated into a module called \maihem \citep{Gray2015,Gray2016} for hydrodynamic simulations with \flash.
The formation of radiatively cooling superwinds and their emission lines has been investigated
using \maihem \citep[][hereafter Paper~I]{Gray2019a}.  In
\citet[][hereafter Paper~II]{Danehkar2021}, we explored the parameter
space for radiatively cooling superwinds using a grid of 
hydrodynamic simulations from \maihem, made in the parameter space of the
metallicity ($Z$), mass loading ($\dot{M}$), wind velocity
($V_{\infty}$), and ambient density ($n_{\rm amb}$), finding
that cooling is enhanced by increasing the gas metallicity and
mass-loading rate, and decreasing the wind velocity.  We employed the
physical conditions produced by our hydrodynamic simulations to create
a grid of collisional ionization and photoionization (CPI) models, 
also predicting UV and optical emission lines under CIE conditions.

Paper~I indicates that NEI states can generate enhanced emission from highly ionized UV lines
such as \ionic{O}{vi} and \ionic{C}{iv}.
Therefore, in the present work, we additionally incorporate NEI states predicted
by \maihem into non-equilibrium photoionization (NPI) models in the
same parameter space grid considered in Paper~II, which allows us to investigate the
implications of NEI calculations for catastrophically cooling superwinds. 
In Section \ref{cooling:winds}, we briefly describe our \maihem hydrodynamic settings and results.
In Section \ref{cooling:photoionization}, we explain how we build NPI modeling with NEI states generated by \maihem. UV emission lines produced by \cloudy and diagnostics diagrams are explained in Section~\ref{cooling:cloudy:results}. Applications of NPI models for observations are discussed in Section~\ref{cooling:observations}, followed by  
a summary and conclusions in Section~\ref{cooling:conclusion}. 

\section{Hydrodynamic Simulations}
\label{cooling:winds}

A complete description of the \maihem code and our adopted parameters is given in Paper~II.  Here, we provide a brief summary.

We use a directionally unsplit hydrodynamic solver \citep{Lee2009,Lee2009a,Lee2013} with a hybrid Riemann solver \citep{Toro1994} and a second-order MUSCL--Hancock reconstruction scheme \citep{vanLeer1979} in the framework of the adaptive mesh hydrodynamics code \flash v4.5 \citep{Fryxell2000} to solve the continuity equation, Euler equation, and energy conservation equation of the fluid model of superwinds with negligible gravitational forces in one-dimensional spherical coordinates that are coupled to the radiative cooling and photo-heating functions of the \maihem package:
\begin{align}
\frac{d \rho _{w}}{d t}+\frac{1}{r^2} \frac{d}{dr} \left( \rho_{w} u_{w} r^2 \right)  &=q_{m},  \label{eq_1} \\
\rho _{w}\frac{d u_{w}}{d t}+\rho_{w} u_{w} \frac{d u_{w}}{dr}+\frac{d P_{w}}{dr}  &=-q_{m} u_{w},  \label{eq_2} \\
\rho _{w}\frac{d {E}_{w}}{d t}+ \frac{1}{r^2} \frac{d}{dr} \bigg[ \rho_{w} u_{w} r^2  \bigg( \frac{u_{w}^{2}}{2} + & \frac{\gamma}{\gamma -1} \frac{P_{w}}{\rho_{w}}  \bigg) \bigg]  \notag \\
&=q_{e}-q_{c}+q_{h},  \label{eq_3}
\end{align}
where $\rho_w$ is the fluid density, ${u}_w$ the fluid velocity, $P_{w}=\left( \gamma -1\right) \rho _{w}\epsilon _{w}$ the thermal pressure with an ideal gas equation of state, ${E}_{w}=\epsilon _{w}+\frac{1}{2}\left\vert u_{w}\right\vert ^{2}$ the total energy per unit mass, $\epsilon _{w}$ the internal energy per unit mass, $\gamma=5/3$ the ratio of specific heats, 
$q_{m} = \dot{M}_{\rm sc} / V_{\rm sc}$ the mass deposition rate per unit volume, 
$q_{e} = \dot{E}_{\rm sc} / V_{\rm sc}$ the energy deposition rate per unit volume, 
$\dot{M}_{\rm sc}$ the mass-loading rate, $\dot{E}_{\rm sc}$ the energy deposition rate,
$V_{\rm sc}= \frac{4}{3} \pi R^3_{\rm sc}$ the SSC volume, $R_{\rm sc}$ the cluster radius,
$q_{c}$ the radiative cooling rates per unit volume, and 
$q_{h}$ the photo-heating rate per unit volume.
In a steady state ($d/d t=0$), the fluid equations take the forms presented in Paper~II that were semianalytically studied by \citet{Silich2004}. Outside the SSC ($r > R_{\rm sc}$), $q_{m}=0$ and $q_{e}=0$, 
while $q_{c}$ and $q_{h}$ also vanish inside the SSC ($r < R_{\rm sc}$) with negligible radiative effects. The steady-state fluid equations reduce to the adiabatic solutions obtained by \citet{Chevalier1985} and \citet{Canto2000} in the absence of the radiative functions $q_{c}$ and $q_{h}$.

As fully described in Paper~II, we set the boundary conditions for the density, temperature, and velocity of the outflow at the cluster boundary ($r=R_{\rm sc}$) according to 
the semianalytic solutions \citep{Chevalier1985,Canto2000,Silich2004} as follows:
\begin{equation}%
\begin{array}
[c]{ccc}%
\rho_{w} = \dfrac{ \dot{M}_{\rm sc}}{ 2 \pi  R_{\rm sc}^2 V_{\infty}},~~~~
& T_{w} =  \dfrac{ 1 }{4\gamma}  \dfrac{\mu_{\rm p}}{ k_{\rm B}} V_{\infty}^2 , ~~~~
u_{w} = \dfrac{1}{2} V_{\infty} ,
\end{array}
\end{equation}
where  $V_{\infty} = ( 2 \dot{E}_{\rm sc} / \dot{M}_{\rm sc} )^{1/2}$ is the actual wind velocity, 
$\mu_{\rm p}=\mu m_{\rm p}$ the mean mass per particle, $\mu \approx 0.61$ the mean atomic weight of particles for 
a fully ionized gas in units of the proton mass $m_{\rm p}$, and $k_{\rm B}$ the Boltzmann constant. 
We also set the initial conditions ($t=0$) of the density, temperature, and velocity of the ambient medium outside the SSC
($r> R_{\rm sc}$) as $\rho_{w} = \mu_{\rm p} n_{\rm amb}$, $T_{w} = T_{\rm amb}$, and $u_{w} = 0$,
where $n_{\rm amb}$ is the number density of the ambient medium, and $T_{\rm amb}$ is the ambient temperature that 
is calculated in PIE by \cloudy for the density profile predicted by an initial \maihem run with $T_{\rm amb}=10^3$\,K. 

\begin{figure}
\center
\includegraphics[width=0.44\textwidth, trim = 0 0 0 0, clip, angle=0]{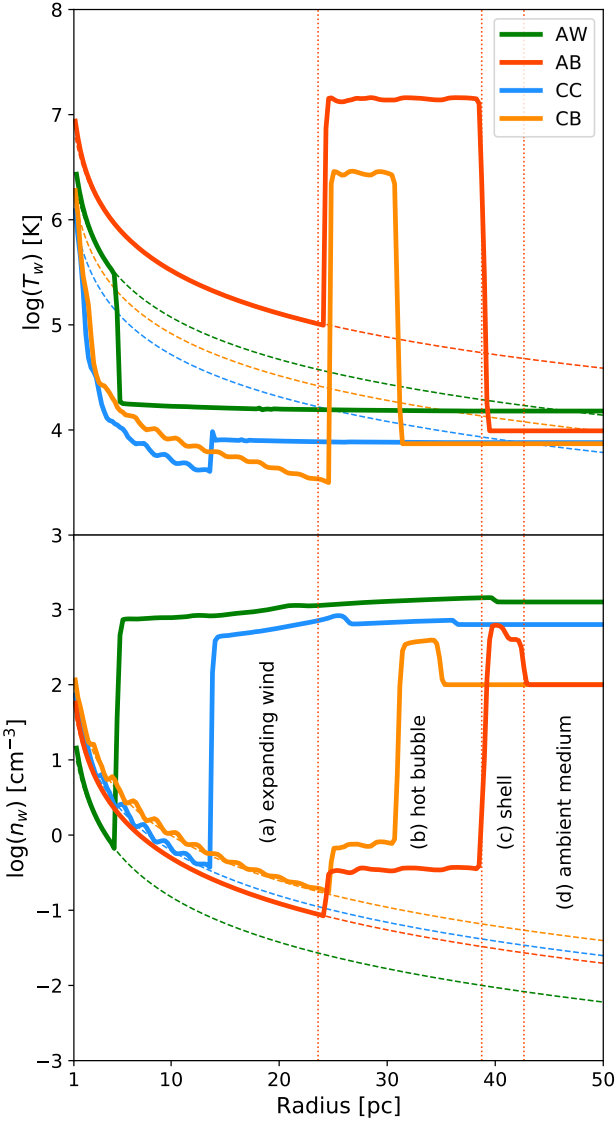}
\caption{\textit{Top Panel}: Wind classification modes based on various temperature profiles, namely the adiabatic wind (AW), adiabatic bubble (AB), catastrophic cooling (CC), and catastrophic cooling bubble (CB), as defined by \citet{Danehkar2021}.
Temperature profiles (solid lines) are plotted along with their adiabatic predictions (dashed lines). 
\textit{Bottom Panel}: The different regions for the AB wind mode are labeled on the density profile and separated by dotted lines: (a) expanding wind, (b) hot bubble, (c) shell, and (d) ambient medium.
Density profiles (solid lines) are plotted along with their adiabatic solutions (dashed lines).
\label{fig:temp:dens:profiles} }
\end{figure}

The radiative cooling rate $q_{c}$ and photo-heating rate $q_{h}$ per volume are calculated by the \maihem module \citep{Gray2019} using 
the ion-by-ion cooling efficiencies $\Lambda_i$, the photo-heating efficiencies $\Gamma_i$, the number densities $n_i$ of ionic species $i$, and the number density of electrons $n_e$:
\begin{equation}%
\begin{array}
[c]{cc}%
q_{c} = \displaystyle\sum_{i}^{}  n_i n_e \Lambda_i ,~~~~
& q_{h} = \displaystyle\sum_{i}^{}  n_i  \Gamma_i,
\end{array}
\end{equation}
where the cooling efficiencies $\Lambda_i$ are calculated for a given wind temperature $T_{w}$ through interpolation on 
the ion-by-ion cooling rates from \citet[][]{Gnat2012} that are also extended down to 5000\,K
by \citet{Gray2015}, and the photo-heating efficiencies $\Gamma_i = \int^{\infty}_{\nu_{0,i}} (4 \pi J_{\nu}/h\nu) h (\nu-\nu_{0,i}) \sigma_{i}(\nu) {\rm d}\nu$ ($\nu$ frequency, $\nu_{0,i}$ ionization frequency, and $h$ the Planck constant) are calculated for a given ionizing spectral energy distribution (SED) $J_{\nu}$ using the photoionization cross sections $\sigma_{i}(\nu)$ from \citet{Verner1995} and \citet{Verner1996} as implemented by \citet{Gray2016} and expanded for further ions in the species network of \citet{Gray2019}. 
The \maihem module for photo-heating calculations is supplied with the ionizing SED made by Starburst99 \citep{Leitherer1999,Leitherer2014} for the fiducial model at age 1\,Myr  with total stellar mass of $M_\star =2.05\times10^6$\,M$_{\odot}$ and metallicity close to the gas metallicity $Z$. 

We produce a grid of hydrodynamic simulations in the parameter space
of the metallicity $Z$, mass loading $\dot{M}_{\rm sc}$, wind velocity $V_{\infty}$, 
and ambient density $n_{\rm amb}$.
We consider the gas metallicity of $\hat{Z} \equiv Z/$Z$_{\odot}=1$, $0.5$, $0.25$, and $0.125$, where $Z/$Z$_{\odot}=1$ associated with the baseline ISM abundances given in Table~2 of Paper~II, 
the mass-loading rate $\dot{M}_{\rm sc} = 10^{-1}$, $10^{-2}$, $10^{-3}$, and $10^{-4} \times \hat{Z}^{0.72}$ M$_{\odot}$\,yr$^{-1}$
according to $ \dot{M}_{\rm sc} \varpropto Z^{0.72}$ \citep{Mokiem2007}, the wind terminal speed $V_{\infty}= 250$, $500$, and $1000 \times \hat{Z}^{0.13}$ km\,s$^{-1}$ according to $V_{\infty} \varpropto Z^{0.13}$ \citep{Vink2001},
and the ambient density $n_{\rm amb}=1$, $10$, $10^2$, and $10^3$\,cm$^{-3}$.
The C/O ratio is also parameterized by the metallicity \citep[described by][]{Danehkar2021}. 
The resulting density and temperature profiles for various parameters are shown as an interactive figure and animation in Figures 2 and 3 of Paper~II.

In Paper~II, we classified our simulated superwinds according to
adiabatic/radiative cooling and the presence/absence of the hot bubble, namely: the adiabatic wind (AW), adiabatic bubble (AB), catastrophic cooling (CC), and catastrophic cooling bubble (CB). Temperature and density profiles of one example of each of the four wind classification modes are plotted in Figure~\ref{fig:temp:dens:profiles}. Moreover, we assigned the adiabatic, pressure-confined (AP) and cooling, pressure-confined (CP) wind modes to those models
whose bubble expansions are stalled by thermal pressures from the
ambient media  (see animation in
Figure~3 of Paper~II).  While models with adiabatic and radiative
cooling (AW and CC) thereby suppress the hot bubble, there are many models
with strong radiative cooling that still do have a bubble (CB). 
Two fully suppressed wind modes were also defined, namely no wind (NW)
and momentum-conserving (MC) evolution.
In the NW mode, the wind is completely inhibited where the supersonic
outflow pressure analytically found by \citet{Canto2000} is less than
the ambient pressure. In the MC mode, radiative cooling caused by high
mass deposition and low heating efficiency is able to completely
suppress the wind before it is launched. The wind classification in the parameter space shown in Figure 4 of Paper~II indicates that radiative cooling is enhanced by an increase in $\dot{M}_{\rm sc}$ and $Z$, and a decrease in $V_{\infty}$.

Our hydrodynamic simulations have been conducted without the gravity module. 
In particular, the Jeans length 
of the ionized ambient medium 
with density $10^3$ cm$^{-3}$ and temperature $10^4$\,K (corresponding to the sound speed  
$\approx 15$ km\,s$^{-1}$), is approximately equal to 100\,pc, which could be comparable to the size of the \ionic{H}{ii} region. Thermal pressure cannot resist gravitational collapse on scales larger than the Jeans length, while 
the self-gravity is negligible below it. 
A gravitational collapse occurs in cool clouds of high density formed by radiative cooling superwinds, which leads to star formation. 
Detailed handling of self-gravity and optical depth for the ambient diffuse radiation is computationally expensive \citep[see e.g.][]{Truelove1997,Wuensch2018,Wuensch2021} and is beyond the scope of this work. 

\section{Non-equilibrium Photoionization Modeling}
\label{cooling:photoionization}

We conduct non-equilibrium photoionization modeling for the physical conditions ($T_{w}$ and $n_{w}$) and  NEI states produced by our \maihem simulations using \cloudy v17.02 \citep{Ferland1998,Ferland2013,Ferland2017}.
The SED and ionizing luminosity of the SSC calculated by Starburst99 are provided as inputs into our \cloudy models to specify an ionizing source (the same as Paper II). 
Non-equilibrium calculations are now made using the outflow temperature $T_w$, outflow density $n_w$ and time-dependent NEI states obtained as a function of the radius $r$ derived from our \maihem  simulations.
In Paper~II, we generated PI models based on the Starburst99 SEDs using only $n_w$, and CPI models with 
$T_w$ and $n_w$ produced by \maihem without the use of NEI states.

Time-dependent calculations of the NEI states are performed by \maihem using the atomic data for recombination, collisional ionization, and photoionization. In non-equilibrium conditions, the number density of each ion $n_{i}$ of each chemical element $A$ evolves according to the time-dependent ionization balance equation \citep[see e.g.][]{Dopita2003}:
\begin{align}
\frac{d n_{A,i}}{d t} = & n_{\rm e} n_{A,i+1}\alpha^{A,i+1}_{\rm rec} - n_{\rm e} n_{A,i}\alpha^{a,i}_{\rm rec}
\notag \\
&   + n_{\rm e} n_{A,i-1} R^{A,i-1}_{\rm coll} - n_{\rm e} n_{A,i}R^{A,i}_{\rm coll}  \notag \\
& +  n_{A,i-1} \zeta^{A,i-1}_{\rm phot} -  n_{A,i} \zeta^{A,i}_{\rm phot},  \label{eq_4}
\end{align}%
where $\alpha^{A,i}_{\rm rec}$ is the recombination coefficient of the ion ${i}$ of element $A$, including the radiative recombination rate \citep{Badnell2006} and dielectronic recombination rate \citep[see Table 1 in][]{Gray2015}, $R^{A,i}_{\rm coll}$ is the collisional ionization rate from \citet{Voronov1997}, $\zeta^{i,A}_{\rm phot}= \int^{\infty}_{\nu_{0,i}} (4 \pi J_{\nu}/h\nu) \sigma_{i, A}(\nu) {\rm d}\nu$ is the photoionization rate of each ion determined from the specified SED field $J_{\nu}$ and the photoionization cross-section $\sigma_{i,A}(\nu)$ \citep{Verner1995,Verner1996}.

\begin{figure}
\begin{center}
\includegraphics[width=0.47\textwidth, trim = 0 0 0 0, clip, angle=0]{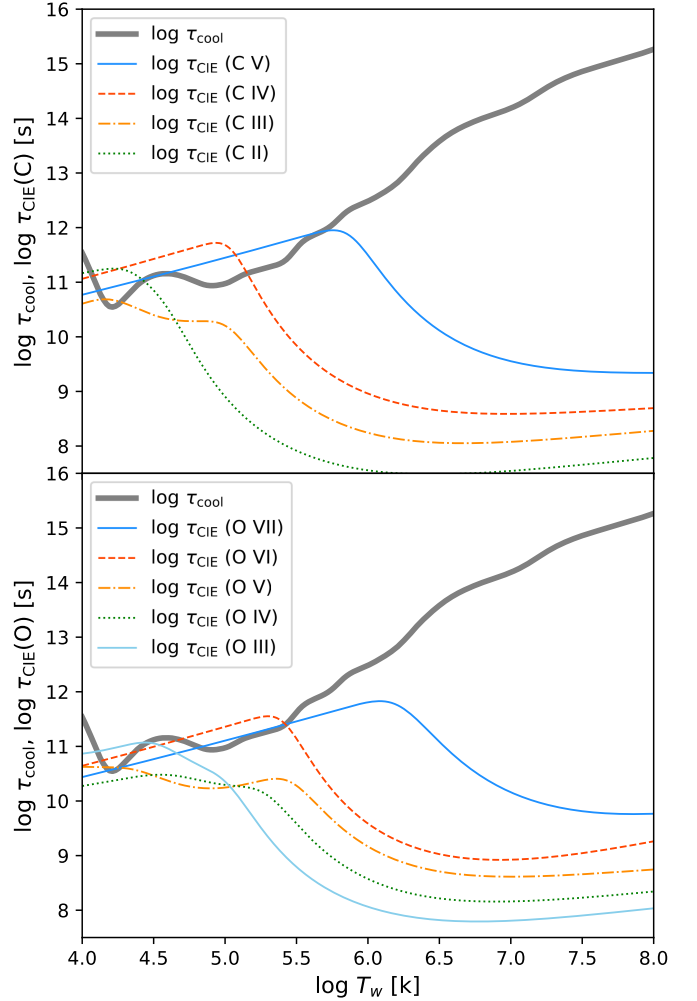}
\end{center}
\caption{The collisional ionization timescales ($\tau_{\rm CIE}$) for ionic species of carbon (top panel) and oxygen (bottom), along with the radiative cooling timescale ($\tau_{\rm cool}$), as a function of the gas temperature ($T_{w}$) for a plasma with $n_{\rm e} =1$\,cm$^{-3}$ and the solar abundances.
\label{fig:cool:time}
}
\end{figure}

Non-equilibrium ionization occurs in regions where the radiative cooling timescale $\tau_{\rm cool} \approx 3 k_{\rm B} T_{w}/ (n_{\rm e} \Lambda)$ is shorter than the collisional ionization timescale $\tau_{\rm CIE} \approx 1 / (n_{\rm e}\alpha^{A,i}_{\rm rec} + n_{\rm e}R^{A,i}_{\rm coll})$, where $\Lambda$ is the total radiative cooling efficiency. 
In the expanding wind region, where density is low ($\lesssim
1$\,cm$^{-3}$), this condition where $\tau_{\rm cool} < \tau_{\rm CIE}$ is
obtained for \ionic{C}{iv} and \ionic{O}{vi} at temperatures below $10^{6}$\,K. 
Thus, these ions may be in NEI states because of radiative cooling, while most ions still remain in CIE $ (\tau_{\rm cool} \gg \tau_{\rm CIE}$).
Figure~\ref{fig:cool:time} shows the radiative cooling timescale and the collisional ionization timescales of different C ions (top panel) and O ions (bottom)  plotted against the electron temperature. 
We calculated the radiative cooling timescale using the total cooling efficiency from \citet[][]{Gnat2012}, and
the CIE timescales using the atomic data for recombination and
collisional ionization.  The figure compares the timescales for $n_{\rm e} =1$\,cm$^{-3}$ and the solar composition. It can be seen that the ionization timescales of \ionic{C}{v}, \ionic{C}{iv}, and \ionic{O}{vi} are longer than the radiative cooling timescale at temperatures below $10^6$\,K, where these ions are in NEI conditions.

To build the NPI models, the density and temperature structures of the outflow extracted from our \maihem simulations
are employed to calculate emissivities by performing \cloudy runs for all individual zones.
The NEI states computed by \maihem with SED are also supplied as inputs to \cloudy. 
The ionizing SED and luminosity produced by Starburst99 are provided
in our \cloudy model to include combined photoionization and
non-equilibrium ionization, decreasing the luminosity by distance from
the SSC as $r^{-2}$.
Following Paper~II, the results predicted by the CPI model are applied to the ambient medium for the ionized, isothermal part of the shell starting outward from $\sim 1$--2\,pc after the interior boundary of the shell.
The final emissivity profiles of emission lines are built by combining
the NEI results of all the outflow zones and  the CPI results of the
ambient medium. The NPI model for the outflow region is therefore
created by zone-by-zone \cloudy calculations based on NEI states
generated by \maihem that requires an individual \cloudy run for each
zone. Typically, there are up to 1024 zones for each simulation.


\begin{figure*}
\begin{interactive}{js}{figure3.tar.gz}
\includegraphics[width=0.8\textwidth, trim = 0 0 0 0, clip, angle=270,origin=rb]{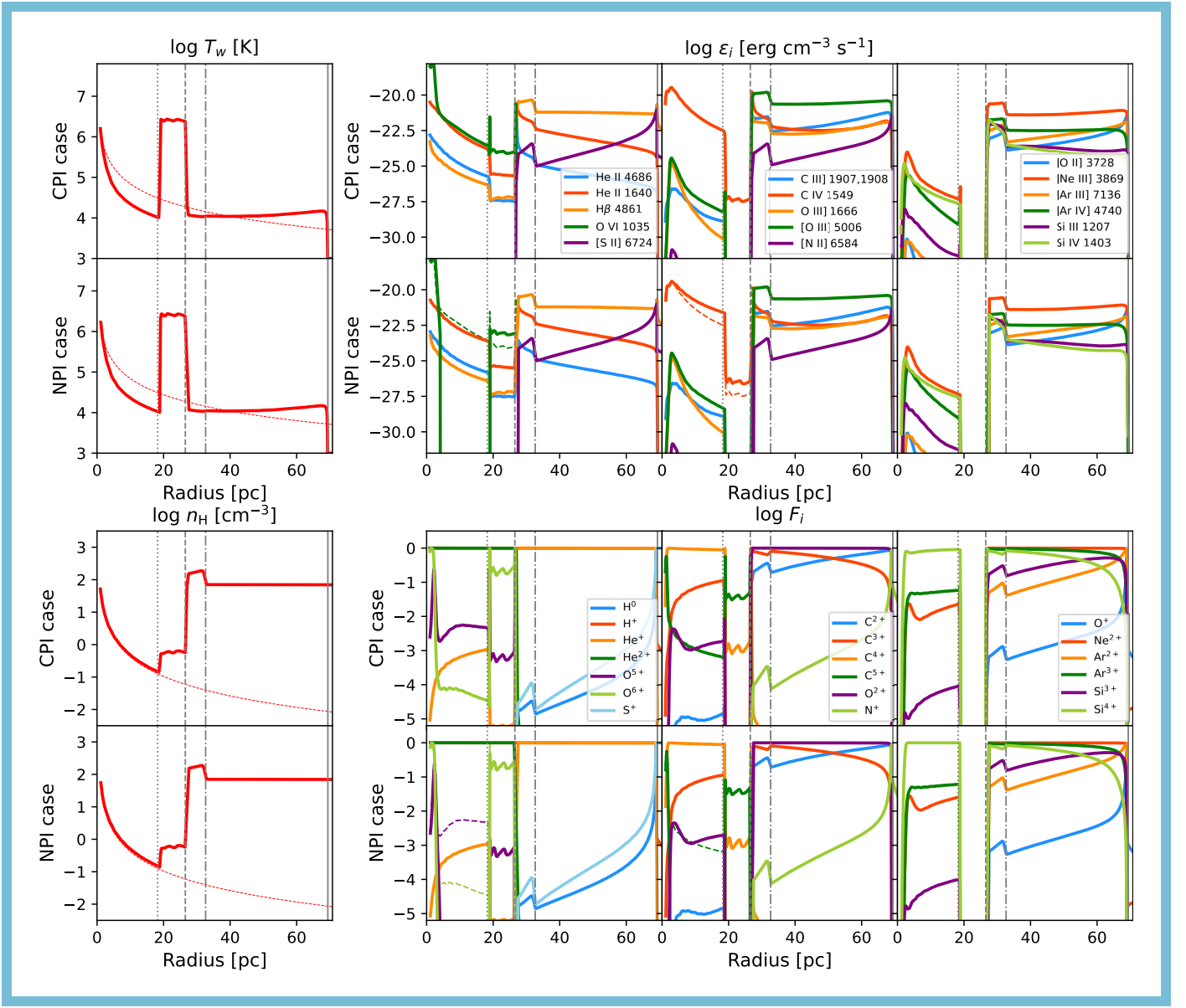}
\end{interactive}
\caption{\textit{Top Panels}: The gas temperature profiles ($T_{w}$\,[K]; left panels) along with the adiabatic prediction (red dashed line), and
  the line emissivities ($\epsilon_{i}$\,[erg\,cm$^{-3}$\,s$^{-1}$]; right
  panels) of the superwind models on a logarithmic scale, from top to
  bottom, in the 
  CPI (collisional ionization $+$ photoionization) cases from \citet{Danehkar2021} and NPI (non-equilibrium photoionization) cases
  (\S\,\ref{cooling:photoionization}), for wind
  speed $V_{\infty}=457$\,km\,s$^{-1}$, metallicity
  $\hat{Z}\equiv Z/$Z$_{\odot}=0.5$, mass-loading rate $\dot{M}_{\rm sc}= 0.607 \times 10^{-2}$
  M$_{\odot}$\,yr$^{-1}$, cluster radius $R_{\rm sc}=1$ pc, and age
  $t=1$ Myr, surrounded by the ambient medium with density $n_{\rm
    amb}=100$ cm$^{-3}$ and temperature $T_{\rm amb}$ determined by
  \cloudy, producing a wind model in the CB mode (described in \S\,\ref{cooling:winds}).
   The \ionic{O}{vi} and \ionf{C}{iv} emissivity profiles predicted by the CPI model are shown by dashed lines in the NPI panel.
   \textit{Bottom Panels}: The hydrogen density profiles ($n_{\rm H}$\,[cm$^{-3}$]; left panels) along with the adiabatic prediction (red dashed lines), and the ionic fractions ($F_{i}$; right panels) on a logarithmic scale for the 
CPI and NPI cases.
  The start and end of the hot bubble (region b), the end of
  the shell (region c), and the Str\"{o}mgren radius are shown by
  dotted, dashed, dash-dotted, and solid lines (gray color),
  respectively. 
  The ionic fractions predicted by the CPI model are plotted by dashed lines in the NPI panel.
  The plots for all the models (192 images) are available in the interactive figure in the online journal. 
  The interactive figure includes menu options to browse all the models by varying $V_{\infty}$, $Z/$Z$_{\odot}$, $\dot{M}_{\rm sc}$, and  $n_{\rm amb}$. 
\label{fig:NEI:emissivity}}
\end{figure*}
 
As described in Paper II, we generate the \cloudy models for the fiducial, $Z/$Z$_{\odot}=1$ model using ISM abundances from \citet{Savage1977} and O/H from \citet{Meyer1998}, along with their depletion factors, and $Z$-parameterized C/O ratio according to the metallicity--C/O correlations from \citet{Garnett1999}. The assumed baseline abundances of elements heavier than helium are also scaled down by \cloudy for $Z/$Z$_{\odot}=0.5$, $0.25$, and $0.125$. In our \cloudy models, we also incorporate typical ISM dust grains with $M_{\rm d}/M_{\rm Z}=0.2$ found by \citet{DeVis2019} for typical galaxies. 
Our \cloudy NPI models similarly use an SED generated by Starburst99 \citep{Levesque2012,Leitherer2014} for a fixed SSC mass of $2.05\times10^6$\,M$_{\odot}$ at all metallicities and age of 1\,Myr using Geneva population grids with stellar rotation \citep{Ekstroem2012,Georgy2012,Georgy2013}, Pauldrach/Hillier atmosphere models \citep{Hillier1998,Pauldrach2001}, and an initial mass function (IMF) with the Salpeter slope $\alpha= 2.35$ for
stellar mass range 0.5--150 M$_{\odot}$. Details of our Starburst99 settings can be found in Paper II, together with the model outputs, including the predicted ionizing luminosity ($L_{\rm ion}$) required for our NPI calculations. 


\section{Emission Line Predictions}
\label{cooling:cloudy:results}

In this section, we describe the volume emissivities of emission lines calculated by \cloudy for the NPI (non-equilibrium photoionization) models, which are used to produce luminosities. Previously, we provided the results for the PI (pure photoionization) and CPI (collisional ionization plus photoionization) models in Paper II. 

For comparison with the CPI models (Paper II), Figure~\ref{fig:NEI:emissivity} presents the emissivities produced for different emission lines as a function of radius for both the CPI and NPI models (top-right panels). 
The CPI model also supplies the NPI model with the emissivity profiles
of the ambient medium that were also calculated using the ambient
temperature structure from the PI model as described in Paper~II.
The emissivities for the outflow region of the NPI model are calculated using the SED ionizing source, hydrodynamic NEI states, and hydrodynamic physical conditions ($T_{w}$ and $n_{w}$).
Figure~\ref{fig:NEI:emissivity} also shows the ionic fractions for both the CPI and NPI models (bottom-right panels). Temperature and density profiles are also presented in Figure~\ref{fig:NEI:emissivity} (left panels). 

In Paper II, we classified our models as optically thin if their ambient media beyond the shell are ionized (H$^+$) in the associated PI models. For the ambient media in the optically thick NPI models, we use the volume emissivity profile from the CPI \cloudy models up to $\sim 2$ pc after the shell inner boundary, where the temperature profile in our hydrodynamic simulations starts to be isothermal.

The luminosity $L_{\lambda}$ of each emission line at wavelength $\lambda$ is determined by taking the integrals of the volume emissivity $\epsilon_{\lambda} (r)$ as follows (see Appendix~A in Paper II):
\begin{equation}
L_{\lambda} =  \int^{2 \pi}_{\varphi=0} \int^{R_{\rm aper}}_{R=0}  \left[ 2 \int^{R_{\rm max}}_{r=R}  \frac{\epsilon_{\lambda} (r)}{ \sqrt{r^2 - R^2} }  r dr \right]  R d R d\varphi,
\label{eq:20}%
\end{equation}
where $r$ is the radial distance of the volume emissivity from the center, $R$ the projected radius from the center, 
$R_{\rm aper}$ the boundary radius of the circular aperture used for the luminosity integration, 
and $R_{\rm max}$ the maximum radius of the boundary in the line of sight. 
We set 
$R_{\rm aper}=R_{\rm max}=R_{\rm Str}$ for radiation-bounded models, 
and $R_{\rm aper}=R_{\rm max}=R_{\rm shell}$ for density-bounded models,  
where $R_{\rm shell}$ is the exterior radius of the shell and 
$R_{\rm Str}$ is the Str\"{o}mgren radius.
As in Paper~II, we also generate partially density-bounded models that
are radiation-bounded in the line of sight, but for a density-bounded aperture:
$R_{\rm aper}=R_{\rm shell}$ and $R_{\rm max}=R_{\rm Str}$.

The emission-line luminosities derived for radiation-bounded,
partially density-bounded, and density-bounded NPI models are listed in Table~\ref{tab:cloudy:output}. The tables for all the NPI model grids are provided in the machine-readable format in Appendix~\ref{appendix:a}. 

\subsection{UV Diagnostic Diagrams}
\label{cooling:opt:diagnostics}

We now generate the UV diagnostic diagrams using the
emission line luminosities obtained from the emissivities predicted by
our NPI models, which include various wind modes such as CC and CB.  
They cover the parameter space of metallicity $Z$, mass-loading $\dot{M}_{\rm sc}$, wind velocity $V_{\infty}$, and ambient density $n_{\rm amb}$.
UV diagnostics diagrams have been used to identify star-forming and active galaxies \citep{Feltre2016,Gutkin2016,Hirschmann2019}.

\begin{figure*}
\begin{center}
\includegraphics[width=0.75\textwidth, trim = 0 0 0 0, clip, angle=0,origin=rb]{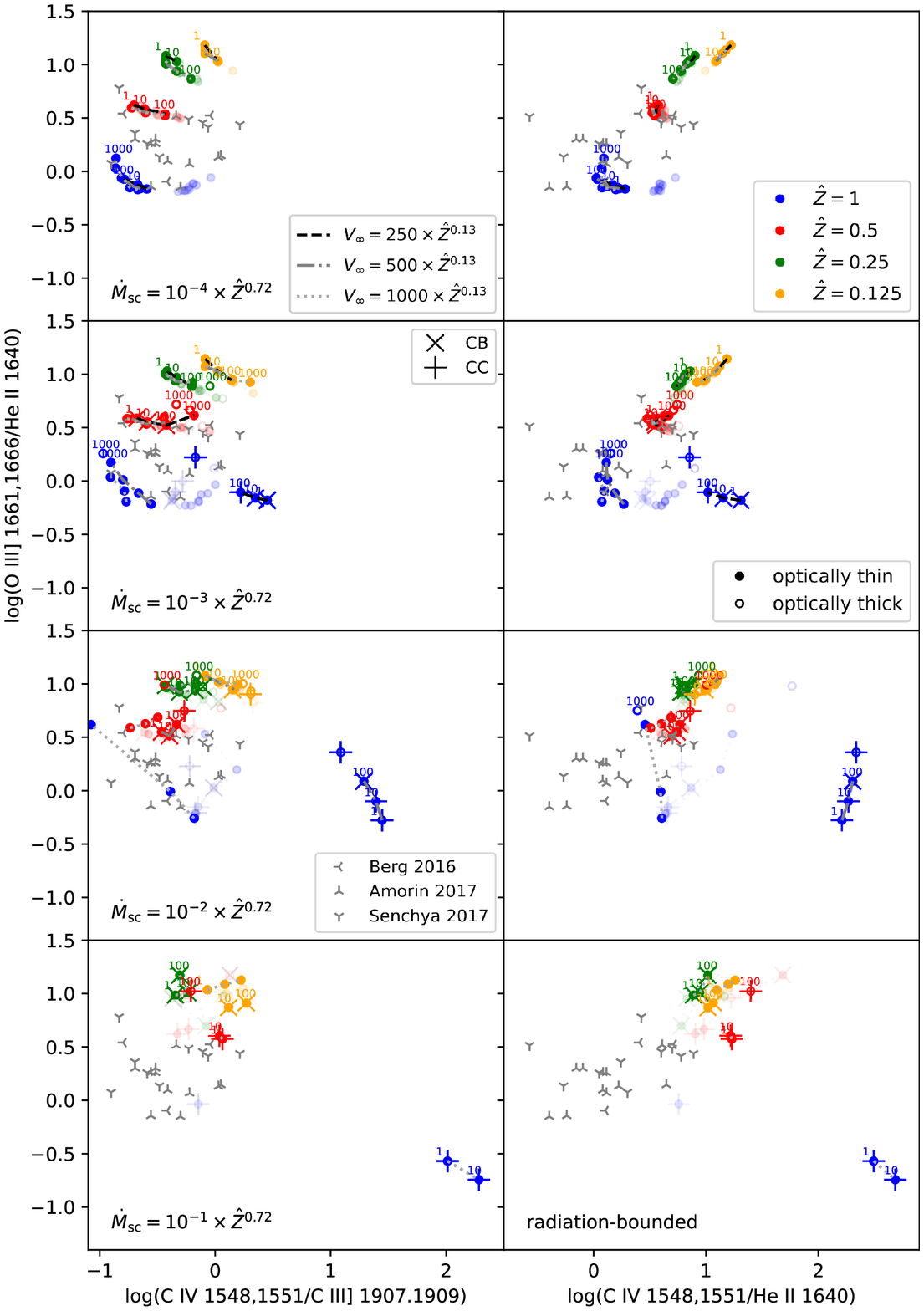}
\end{center}
\caption{The UV diagnostic diagrams plotted \ionf{O}{iii}] $\lambda\lambda$1661,1666/\heii $\lambda$1640 versus \ionic{C}{iv} $\lambda\lambda$1548,1551/\ionf{C}{iii}] $\lambda\lambda$1907,1909 (left) and  \ionf{O}{iii}] $\lambda\lambda$1661,1666/\heii $\lambda$1640 versus \ionic{C}{iv} $\lambda\lambda$1548,1551/\heii $\lambda$1640 (right panels) for the fully radiation-bounded NPI models with mass-loading rates $\log \dot{M}_{\rm sc} =-4$, $-3$, $-2$, and $-1$ M$_{\odot}/$yr (from top to bottom), ambient densities $n_{\rm amb}= 1$, 10, $10^2$, and $10^3$\,cm$^{-3}$ (labeled on plots), metallicities $\hat{Z} \equiv Z/$Z$_{\odot}=1$ (blue), $0.5$ (red), $0.25$ (green), and $0.125$ (yellow color), and wind velocities $V_{\infty}=250$ (dashed), $500$ (dash-dotted), and $1000$ km\,s$^{-1}$ (dotted lines). For the sub-solar models, we use the solar model parameters scaled as $\dot{M}_{\rm sc} \varpropto Z^{0.72}$ and $V_{\infty} \varpropto Z^{0.13}$. The optically thin and thick models are plotted by filled and empty circles, respectively. 
The wind catastrophic cooling (CC) and catastrophic cooling with the bubble (CB) modes are labeled by the plus ('$+$') and cross ('$\times$') symbols, respectively.
The CPI results from Paper II are shown by lightly shaded colors. 
The plotted observations are from \citet{Berg2016}, \citet{Amorin2017}, and \citet{Senchyna2017}.
\label{fig:uv:diag:radi}
}
\end{figure*}

\begin{figure*}
\begin{center}
\includegraphics[width=0.75\textwidth, trim = 0 0 0 0, clip, angle=0,origin=rb]{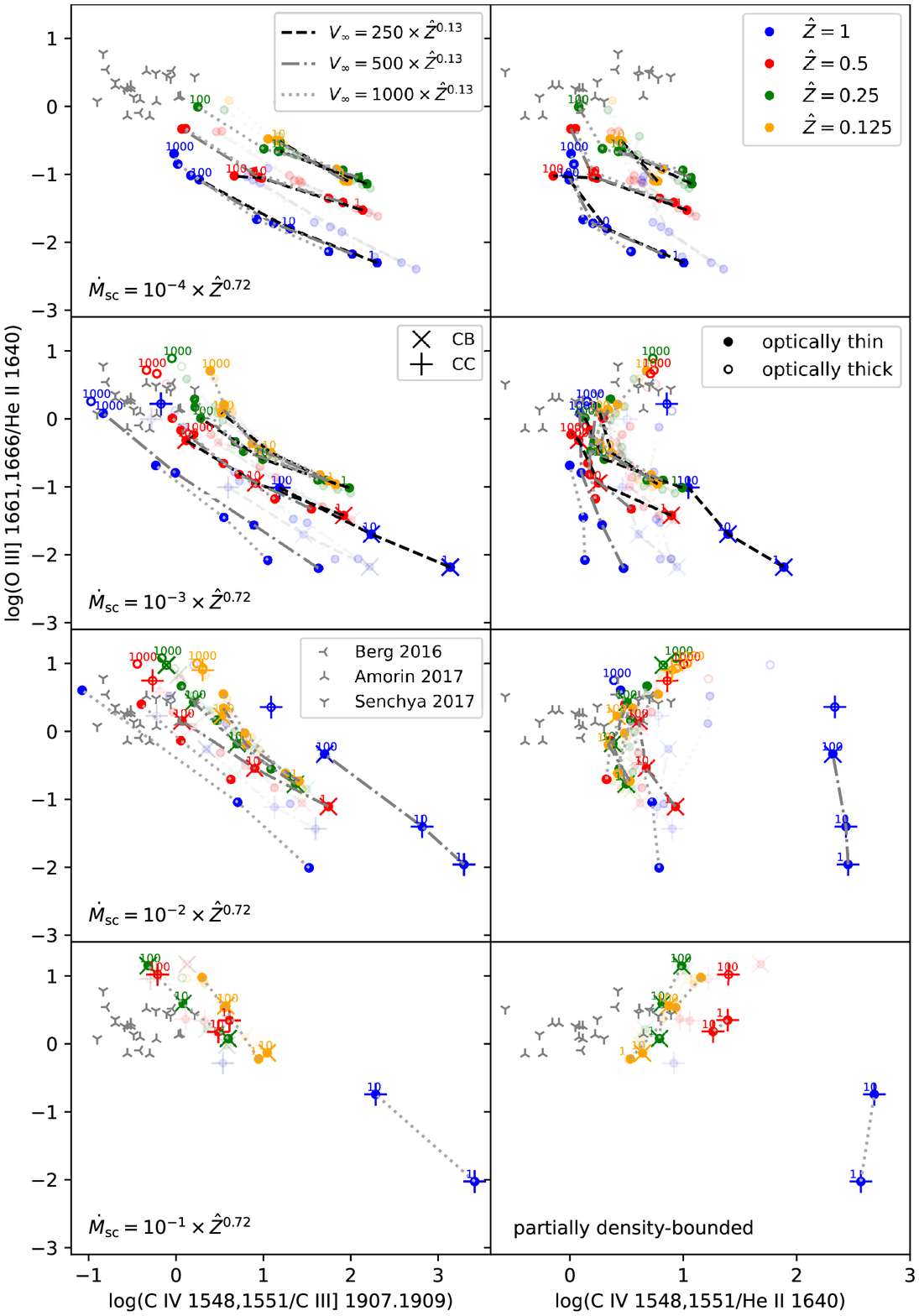}
\end{center}
\caption{The same as Figure~\ref{fig:uv:diag:radi}, but for the partially density-bounded models.
\label{fig:uv:diag:pden} 
}
\end{figure*}

Figures~\ref{fig:uv:diag:radi} and \ref{fig:uv:diag:pden} show \ionf{O}{iii}] $\lambda\lambda$1661,1666/\heii $\lambda$1640 versus \ionic{C}{iv} $\lambda\lambda$1548,1551/\ionf{C}{iii}]
$\lambda\lambda$1907,1909 and \ionic{C}{iv} $\lambda\lambda$1548,1551/\heii $\lambda$1640, for our
radiation-bounded and partially density-bounded models, respectively. 
As seen in Figure~\ref{fig:uv:diag:radi}, the \ionf{O}{iii}]/\heii and  \ionic{C}{iv}/\ionf{C}{iii}] ratios decrease with an increase in the metallicity.
However, the radiation-bounded models with $Z/$Z$_{\odot}=1$ that experience strong radiative cooling produce higher values of the \ionic{C}{iv}/\ionf{C}{iii}] ratio.
Moreover, we notice some slight enhancements in the \ionic{C}{iv}/\ionf{C}{iii}] ratio in the models with $Z/$Z$_{\odot}=0.5$ and $\dot{M}_{\rm sc}$\,$\geqslant $\,$10^{-2} \hat{Z}^{0.72}$\,M$_{\odot}/$yr, where strong radiative cooling occurs. 
Cooling produces strong \ionic{C}{iv} emission within the
free-expanding wind region, displacing some models relative to adiabatic ones.  
For the models with lower mass-loading rates ($\dot{M}_{\rm sc}$\,$\leqslant $\,$10^{-3} \hat{Z}^{0.72}$\,M$_{\odot}/$yr) 
and lower metallicity ($Z/$Z$_{\odot}\leqslant0.5$), radiative cooling
is not strong enough to generate significant \ionic{C}{iv} emission. 
In the partially density-bounded models having radiatively cooled
winds (Figure~\ref{fig:uv:diag:pden}), enhanced \ionic{C}{iv} emission
is more pronounced, since these models are more strongly weighted toward emission produced in the free wind and the shell.
On the other hand,
\ionf{O}{iii}] and \ionf{C}{iii}] are less sensitive to strong radiative cooling, so they do not demonstrate any significant departures from the adiabatic wind models.

\begin{figure}
\begin{center}
\includegraphics[width=0.42\textwidth, trim = 0 0 0 0, clip, angle=0,origin=rb]{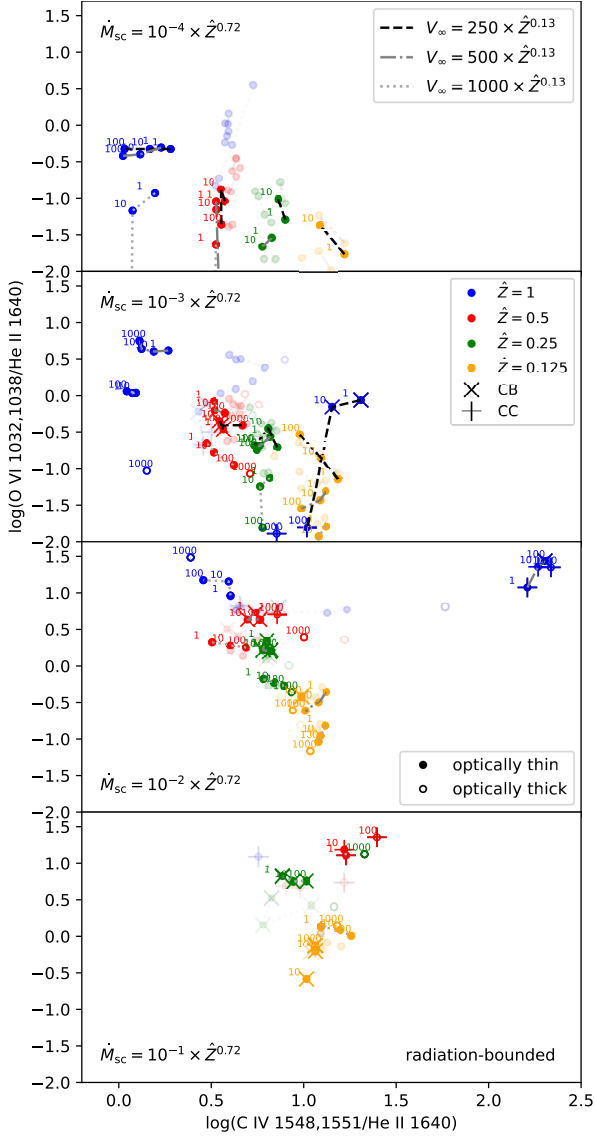}
\end{center}
\caption{
The UV diagnostic diagrams plotted \ovi $\lambda\lambda$1032,1038/\heii $\lambda$1640 versus \ionic{C}{iv} $\lambda\lambda$1548,1551/\heii $\lambda$1640 for the radiation-bounded NPI models. 
Symbols and line types are as in Figure~\ref{fig:uv:diag:radi}. The CPI results are shown by lightly shaded colors.
\label{fig:uv3:diag:radi}
}
\end{figure}

\begin{figure}
\begin{center}
\includegraphics[width=0.42\textwidth, trim = 0 0 0 0, clip, angle=0,origin=rb]{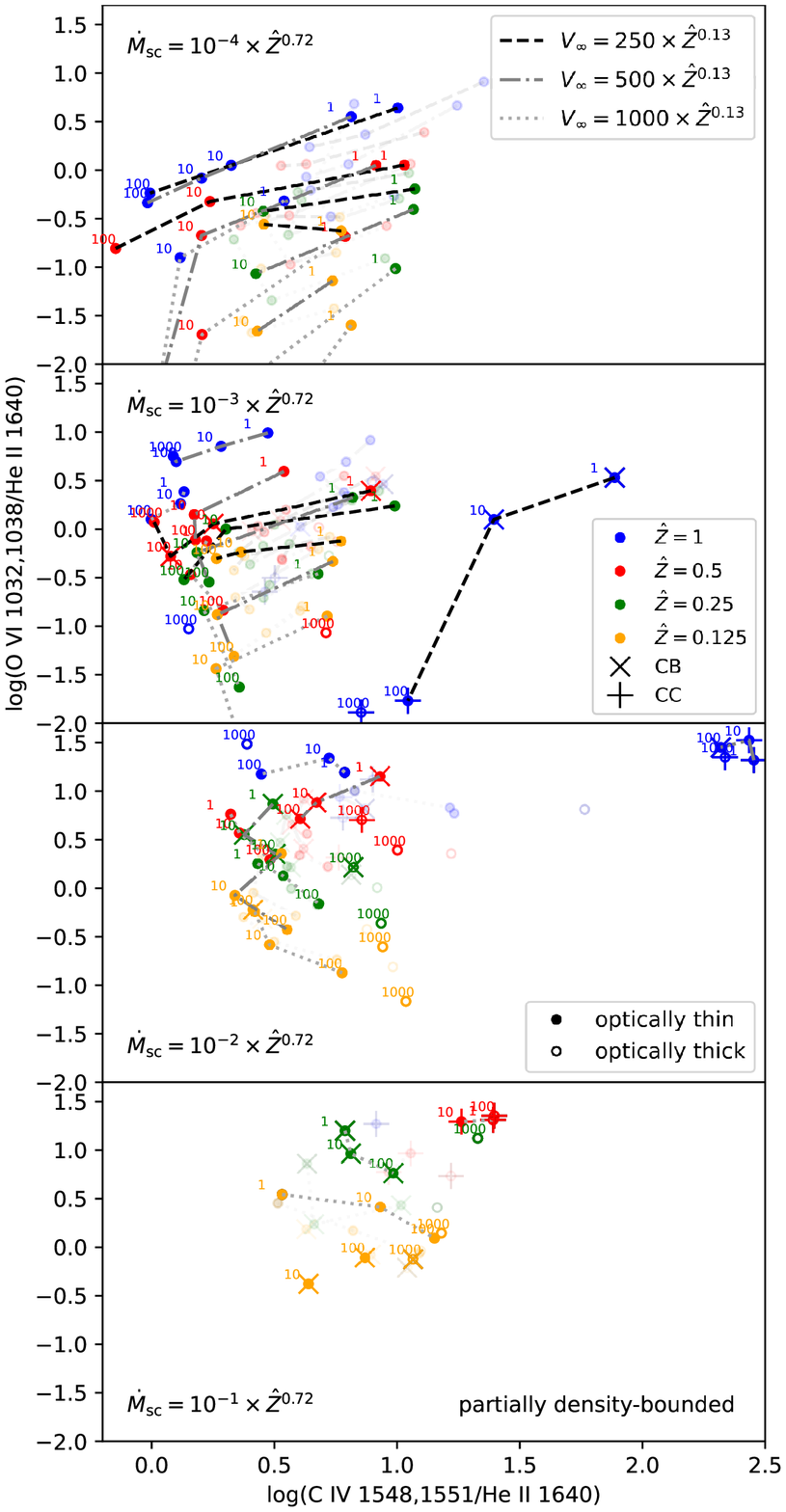}
\end{center}
\caption{The same as Figure~\ref{fig:uv3:diag:radi}, but for the partially density-bounded models.
\label{fig:uv3:diag:pden} 
}
\end{figure}

Similarly, we present diagnostic diagrams for
\ovi$\lambda\lambda$1032,1038/\heii $\lambda$1640 versus \civ
$\lambda\lambda$1549,1551/\heii $\lambda$1640 in Figures~\ref{fig:uv3:diag:radi}
and \ref{fig:uv3:diag:pden} for radiation-bounded, and partially
density-bounded, respectively, which may be compared to those in Paper~II (CPI results plotted by lightly shaded colors).
The highly ionized \ovi $\lambda\lambda$1032,1038 emission has been suggested as a
diagnostic tool for catastrophic cooling winds \citep{Gray2019}.
At first glance, it seems that the  \ionic{C}{iv} $\lambda\lambda$1548,1551 emission rather \ovi $\lambda\lambda$1032,1038
is significantly enhanced in the models with catastrophic cooling.
However, we also see in Figure~\ref{fig:uv3:diag:radi} that strong radiative cooling considerably increases the \ovi emission in 
the models with $\dot{M}_{\rm sc}$\,$= $\,$10^{-2}$ and $10^{-3} \times \hat{Z}^{0.72}$\,M$_{\odot}/$yr at $Z/$Z$_{\odot}= 0.5$ and $0.25$, 
while the \civ emission from the radiatively cooled winds is slightly
increased in the models with $\dot{M}_{\rm sc}$\,$=
$\,$10^{-2}\hat{Z}^{0.72}$\,M$_{\odot}/$yr, remaining
similar to that from the adiabatic models for $\dot{M}_{\rm sc}$\,$= $\,$10^{-3}\hat{Z}^{0.72}$\,M$_{\odot}/$yr.
While the \civ/\heii emission-line ratio expected from the NPI models may be employed to distinguish between radiatively cooling and adiabatic winds in metal-rich ($Z/$Z$_{\odot}\gtrsim 0.5$) regions, the non-equilibrium photoionized \ovi/\heii emission-line ratio could be used as a diagnostic of radiative cooling in metal-poor ($Z/$Z$_{\odot}\lesssim 0.5$) \hii\ regions typical of starburst galaxies.

Previously, we also saw enhancement of radiation-bounded \ovi emission predicted by the CPI models
for radiatively cooled outflows with mass-loading rates $\geqslant
10^{-2}$\,M$_{\odot}/$yr in metal-poor regions
$Z/$Z$_{\odot}\leqslant0.5$ (Figure\,11 in Paper~II).
The enhanced \ovi/\heii emission-line ratios produced by
non-equilibrium photoionization processes in metal-poor \hii\ regions
is similar to that seen in combined collisional ionization and photoionization processes, 
whereas the substantially enhanced \civ/\heii ratios in the NPI models
are not seen in the CPI 
models presented in Paper II. As seen in Figure~\ref{fig:NEI:emissivity}, the \ovi\ emission is mostly generated in the central part of the free wind close to the SSC and in the hot bubble region in our NPI models. 
The \ovi\ emissivity profile for NPI is different from that for the
CPI model (see Figure~\ref{fig:NEI:emissivity}), while the enhanced
\civ\ emissivity profile in NPI looks roughly similar to the CPI
model, except for some sharp peaks in \ovi\ and \civ\ in the CPI model
in the interface region between the bubble (CIE) and the shell (PIE).

It can be seen in Figures~\ref{fig:uv3:diag:radi} and \ref{fig:uv3:diag:pden} that
the \ovi\ emission from the adiabatic winds rises with increasing metallicity and mass deposition.
This trend in the adiabatic winds of the NPI models that is similar to those of the CPI models in Paper II
is more noticeable in the partially density-bounded models in Figure~\ref{fig:uv3:diag:pden}.
The enhanced \ovi\ emission at $Z/$Z$_{\odot}<1$ agrees with Paper~I suggesting that \ovi\ could be
enhanced in radiatively cooled winds, although not significantly for $Z/$Z$_{\odot}= 1$.
However, the \civ\ emission appears to be a useful diagnostic of radiative cooling at $Z/$Z$_{\odot}= 1$.
Thus, to identify strong radiative cooling,  it is necessary to know the metallicity.
Moreover, information on the ambient density and mass loading further help to diagnose
radiatively cooling superwinds.

Figures~\ref{fig:uv:diag:radi}--\ref{fig:uv3:diag:pden} also include optically thick NPI models. 
In particular, the enhanced \ovi\ and \civ\ emission are also seen in some of the optically thick NPI models with strong radiative cooling. 
While \civ\ emission did not increase much in the CPI model in Paper II, 
substantially enhanced \civ\ emission is seen in the radiatively cooling wind models with $Z/$Z$_{\odot}= 1$, as well as those with $Z/$Z$_{\odot}= 0.5$ and $\dot{M}_{\rm sc} \geqslant 10^{-2}$\,M$_{\odot}/$yr.

We note that our models do not account for radiative
transfer, in particular, absorption and reemission by dust and other species.
We incorporate metal depletion into our hydrodynamic simulations and 
include dust grains and depletion in our \cloudy calculations using the
NEI states and physical conditions produced by \maihem.   
Including dust grains as separate species with 
distinctive equations of state in \maihem hydrodynamic simulations
would contribute to significant changes in the UV diagnostic diagrams, 
since dust grains can affect thermal structures and absorb radiation fields. 
Moreover, NEI cooling rates can also be affected by dust grains 
\citep[][]{Richings2014}.
Currently, we assume typical ISM dust grains with $M_{\rm d}/M_{\rm Z}=0.2$, in typical galaxies \citep{DeVis2019}.

\vfill\break

\section{Applications for Observations}
\label{cooling:observations}

Our NPI models predict that radiatively cooling outflows could contribute to
the higher excitation seen in extreme starbursts and GPs.
In Figures~\ref{fig:uv:diag:radi}--\ref{fig:uv:diag:pden}, we plot 
UV observations of nearby and distant starburst galaxies
\citep{Richard2011,Masters2014,Berg2016,Amorin2017,Senchyna2017}.
Their physical properties are roughly similar to those considered for our models.
The nearby dwarf galaxies analyzed by  \citet{Berg2016} and \citet{Senchyna2017} have 
oxygen abundances of $12+\log($O/H$)\approx 7.6$ and $\lesssim 8.3$, respectively;
and the more distant star-forming galaxies studied by \citet{Amorin2017} have
a mean oxygen abundance of $12+\log($O/H$) \approx 7.6$.

In particular, our models show that
\ionic{O}{vi} $\lambda\lambda$1032,1038 emission is
produced in radiatively cooling outflows, as suggested by
\citet{Heckman2001}, \citet{Otte2003}, \citet{Hayes2016}, and \citet{Li2017}.
As seen in Figures~\ref{fig:uv3:diag:radi}--\ref{fig:uv3:diag:pden}, 
particularly in the metal-poor ($Z/$Z$_{\odot}\leqslant 0.5$) models, 
the \ionic{O}{vi} $\lambda\lambda$1032,1038 emission lines predicted by the radiative cooling wind models are higher  
than those expected by the adiabatic wind models with the same metallicity and mass loading.

Strong \ionic{O}{vi} emission indeed is sometimes seen in young starbursts.
For example, \citet{Marques-Chaves2021} find clear P-Cygni emission in \ionic{O}{vi},
\ionic{Si}{iv}, and \ionic{C}{iv}, in a luminous LyC emitter
at $z=3.2$.  The strength of these features is hard to
explain with a simple starburst stellar population.  While the broad
line profiles must be dominated by stars, it may be possible that a
component from radiatively cooling outflows could contribute.
Similarly, \ionic{O}{vi} $\lambda\lambda$1032,1038 emission lines with
P Cygni profiles were also identified in Haro\,11
\citep{Bergvall2006,Grimes2007} and other starbursting blue compact galaxies \citep{Izotov2018a}.
\ionic{O}{vi} emission was detected by \citet{Otte2003}
toward a soft X-ray bubble in the star-forming galaxy NGC\,4631 that
could be associated with cooling, galactic outflows ($v_{\rm w} \sim 50$--100\,km\,s$^{-1}$).
More recently, \ionic{O}{vi} imaging of the starburst SDSS J1156$+$5008 by \citet{Hayes2016}
shows an extended halo, and its spectrum also contains \ionic{O}{vi} absorption outflowing
with an average velocity of $380$\,km\,s$^{-1}$.
Observations of the dwarf starburst galaxy NGC\,1705 studied by \citet{Heckman2001}
revealed a low-speed outflow ($v_{\rm w} = 77$\,km\,s$^{-1}$) in \ionic{O}{vi} absorption,
surrounded by a $\sim 10^4$\,K medium 
arising in a conductive interface between the hot superbubble and cool outer shell, 
and which cannot be explained by a simple adiabatic superbubble model.
The physical properties of these starburst galaxies
are in the parameter ranges used by our models.

Our models with catastrophically cooling winds at $Z/$Z$_{\odot}\geqslant 0.5$ also
demonstrate prominent \ionic{C}{iv} emission lines relative to
adiabatic wind models.
Interestingly, \citet{Senchyna2017} found 
strong \civ\ emission
in metal-poor star-forming regions associated with
minimal stellar wind features and $12+\log
($O/H$)\lesssim 8.3$ ($Z/$Z$_{\odot} \lesssim 0.6$). 
Similarly, \citet{Berg2019a} detected intense \ionic{C}{iv}
$\lambda\lambda$1548,1551 emission lines from
two extreme UV emission-line galaxies demonstrating little-to-no
outflows \citep{Berg2019a} at $12+\log ($O/H$) \approx 7.5$ ($Z/$Z$_{\odot} \sim 0.1$), as well as 
other nearby metal-poor high-ionization dwarf galaxies \citep{Berg2019b}. 
Our non-equilibrium photoionization models predict that such intense \civ emission could be prevalent in
radiatively cooling winds with $Z/$Z$_{\odot}\gtrsim 0.5$, but if these
objects have C-enhanced abundances, radiatively cooling \ionic{C}{iv}
could contribute to the unusually strong emission at lower metallicities.
With constraints on metallicity and mass loading, it is becomes possible to
distinguish between adiabatic and radiatively cooling outflows.

\section{Summary and Conclusions}
\label{cooling:conclusion}

We have presented here our grid of non-equilibrium photoionization (NPI) models 
constructed using non-equilibrium ionization (NEI) states and physical conditions 
predicted by our hydrodynamic simulations previously
presented in Paper II. 
We use the same ionizing SED associated with a total stellar mass of $2.05\times10^6$\,M$_{\odot}$,
and the same parameter space of the metallicity ($\hat{Z} \equiv Z/$Z$_{\odot} = 1$, $0.5$, $0.25$, $0.125$), mass-loading rate ($\dot{M}_{\rm sc}$\,$=$\,$10^{-1},\ldots,10^{-4} \times \hat{Z}^{0.72} $\,M$_{\odot}\,$yr$^{-1}$), 
wind velocity ($V_{\infty}$\,$=$\,$250,500,1000 \times \hat{Z}^{0.13} $\,km\,s$^{-1}$), and
ambient density ($n_{\rm amb}$\,$=$\,$1,\ldots,10^3$\,cm$^{-3}$) employed in Paper II.
The non-equilibrium photoionization modeling is carried out using 
the same non-CIE approach implemented by \citet{Gray2019a}.

Previously, we identified the parameter space (in $\hat{Z}$, $\dot{M}_{\rm sc}$, $V_{\infty}$, $n_{\rm amb}$) associated with strongly radiative cooling, and classified them under different
wind modes based on departures from the adiabatic solutions and
presence/absence of the hot bubble within 1\,Myr
\citep[see \S\,4 in][]{Danehkar2021}.
We found that low  heating efficiency and high
mass deposition are linked to strong radiative cooling effects,
while the presence of a hot bubble is not a reliable indicator of either adiabatic or 
radiatively cool outflows.
We also employed physical conditions to predict
emission lines for combined collisional ionization 
and photoionization (CPI) processes without considering 
non-equilibrium ionization conditions.

In this paper, we utilize NEI states and physical proprieties 
generated by time-dependent non-equilibrium processes
to calculate volume emissivities 
of UV and optical emission lines with \cloudy, and their luminosities for
radiation-bounded and partially density-bounded models.
We use the predicted line luminosities of UV lines studied in Paper~II to compile diagnostic diagrams 
for comparison with observations of star-forming regions. 

The radiation-bounded UV emission-line ratios predicted by our NPI
models are generally located
in our diagnostic diagrams where 
both nearby and distant starburst galaxies 
with the modeled physical properties are observed
(see Figures~\ref{fig:uv:diag:radi}--\ref{fig:uv:diag:pden}), similar
to what is seen for the CPI models in Paper II.
However, we see some noticeable differences between the 
emission-line ratios made by our NPI models and those by the CPI models
that are due to NEI conditions.
Our NPI models demonstrate that radiative cooling strongly enhances
\ionic{C}{iv} emission in metal-rich ($Z/$Z$_{\odot}\geqslant 0.5$) regions 
and also increases \ovi\ emission in metal-poor ($Z/$Z$_{\odot}\leqslant 0.5$) regions that are 
typical of starbursts.
However, some constraints on the metallicity and mass loading are necessary in order to
diagnose superwinds with strongly radiative cooling.

The enhanced \ionic{C}{iv} $\lambda\lambda$1548,1551 and \ovi\ $\lambda\lambda$1032,1038 in catastrophically cooling outflows
were previously suggested based on non-equilibrium ionization models by 
\citet{Gray2019a} and \citet{Gray2019}.
Moreover, hydrodynamic simulations of starburst-driven galactic outflows by \citet{Cottle2018} could not adequately 
produce the observed \ionic{O}{vi}, implying possible implications of nonequilibrium ionization processes.
Previously, time-dependent calculations of the NEI states by \citet{deAvillez2012} 
showed the production of \ovi\ in the thermally stable ($\sim 10^{\textrm{3.9--4.2}}$) and unstable ($\sim 10^{\textrm{4.2--5}}$) regimes having temperatures below that required for \ovi\ in collisional ionization.
Our photoionization calculations done with NEI states demonstrate the feasibility of 
intense \ovi\ made by radiative cooling in metal-poor environments.

The occurrence of catastrophic cooling and the formation of \ionf{C}{iv} and \ovi\ emission should also be 
investigated under different physical conditions such as 
compact and ultra-compact \hii\ regions where cluster dimensions are much smaller
than our typical SSC radius (1\,pc) and
stellar masses are lower than our SSC mass. 
It is also necessary to explore different dust proprieties
that are typically seen in distant star-forming galaxies.
Future studies on the wide range of physical parameters will help us
broaden our understanding of radiatively cooling outflows and their 
associated emission lines.

\begin{acknowledgments}

We are grateful to Richard W\"{u}nsch for careful review of the manuscript. The hydrodynamics code \flash used in this work was developed in part by the DOE NNSA ASC- and DOE Office of Science ASCR-supported Flash Center for Computational
Science at the University of Chicago.
Analysis and visualization of the \flash simulation data were 
performed using the yt analysis package \citep{Turk2011}.

\end{acknowledgments}

\begin{appendix}

\section{Supplementary Material}
\label{appendix:a}

The interactive figure (192 images) of Figure~\ref{fig:NEI:emissivity} is available in the electronic edition of this article, and is archived on Zenodo (doi:\href{https://doi.org/10.5281/zenodo.6601127}{10.5281/zenodo.6601127}).
This interactive figure is also hosted at:  \url{https://galacticwinds.github.io/superwinds}.

The 3 machine-readable tables with the emission-line data (including Table~\ref{tab:cloudy:output}) is available in the electronic edition of this article. 
Each file is named as \verb|table_case_bound.dat|, such as
\verb|table_NPI_radi.dat|, where \verb|case| is for the
ionization case (\verb|NPI|: non-equilibrium photoionization), and \verb|bound| for the
optical depth model (\verb|radi|: fully radiation-bounded, \verb|pden|: partially density-bounded, and \verb|dens|: fully density-bounded). Each file contains the following information:

\begin{table}
\begin{center}
\caption[]{Emission line luminosities calculated by the NPI models on a logarithmic scale (unit in erg/s) with different optical depth configurations (see Appendix~\ref{appendix:a} for more information). 
\label{tab:cloudy:output}}
\scriptsize
\begin{tabular}{l|c|c|c}
  \hline\hline\noalign{\smallskip}
\multicolumn{1}{l|}{Emission Line} & \multicolumn{1}{c|}{radiation-bound} & \multicolumn{1}{c|}{part.density-bound} & \multicolumn{1}{c}{density-bound} \\
\noalign{\smallskip}
\tableline
\noalign{\smallskip}
Ly$\alpha$     $\lambda$1216 &  40.561 &  40.151 &  40.025 \\
H$\alpha$      $\lambda$6563 &  40.791 &  40.463 &  40.186 \\
H$\beta$      $\lambda$4861 &  40.352 &  40.024 &  39.748 \\
\ionic{He}{i}     $\lambda$5876 &  39.443 &  39.118 &  38.843 \\
\ionic{He}{i}     $\lambda$6678 &  38.876 &  38.550 &  38.275 \\
\ionic{He}{i}     $\lambda$7065 &  39.121 &  38.818 &  38.532 \\
\ionic{He}{ii}    $\lambda$1640 &  38.767 &  38.685 &  38.588 \\
\ionic{He}{ii}    $\lambda$4686 &  36.311 &  36.238 &  36.153 \\
\ionic{C}{ii}     $\lambda$1335 &  38.402 &  37.941 &  37.546 \\
\ionic{C}{ii}     $\lambda$2326 &  38.177 &  37.360 &  35.930 \\
\ionic{C}{iii}    $\lambda$977  &  38.574 &  38.347 &  38.117 \\
\ionf{C}{iii}]   $\lambda$1909 &  39.878 &  39.261 &  38.669 \\
\ionic{C}{iii}    $\lambda$1549 &  36.902 &  36.679 &  36.451 \\
\ionic{C}{iv}     $\lambda$1549 &  39.529 &  39.290 &  39.177 \\
{[\ionf{N}{i}]}      $\lambda$5200 &  36.092 &  35.172 &  31.005 \\
{[\ionf{N}{ii}]}     $\lambda$5755 &  37.051 &  36.207 &  34.730 \\
{[\ionf{N}{ii}]}     $\lambda$6548 &  38.179 &  37.337 &  35.992 \\
{[\ionf{N}{ii}]}     $\lambda$6583 &  38.649 &  37.806 &  36.462 \\
\ionic{N}{iii}    $\lambda$1750 &  39.343 &  38.679 &  37.966 \\
\ionic{N}{iii}    $\lambda$991 &  39.062 &  38.820 &  38.584 \\
\ionic{N}{iv}     $\lambda$1486 &  39.173 &  38.774 &  38.453 \\
\ionic{N}{v}      $\lambda$1240 &  39.593 &  39.593 &  39.593 \\
\ionic{O}{i}      $\lambda$1304 &  38.011 &  37.092 &  33.792 \\
{[\ionf{O}{i}]}      $\lambda$6300 &  36.234 &  35.323 &  31.558 \\
{[\ionf{O}{i}]}      $\lambda$6364 &  35.739 &  34.827 &  31.063 \\
{[\ionf{O}{ii}]}     $\lambda$3726 &  38.609 &  37.842 &  36.955 \\
{[\ionf{O}{ii}]}     $\lambda$3729 &  38.753 &  37.981 &  37.055 \\
{[\ionf{O}{ii}]}     $\lambda$7323 &  37.364 &  36.815 &  36.385 \\
{[\ionf{O}{ii}]}     $\lambda$7332 &  37.281 &  36.729 &  36.296 \\
{\ionf{O}{iii}]}    $\lambda$1661 &  38.791 &  38.244 &  37.778 \\
{\ionf{O}{iii}]}    $\lambda$1666 &  39.262 &  38.715 &  38.250 \\
{[\ionf{O}{iii}]}    $\lambda$2321 &  38.549 &  38.066 &  37.670 \\
{[\ionf{O}{iii}]}    $\lambda$4363 &  39.148 &  38.665 &  38.270 \\
{[\ionf{O}{iii}]}    $\lambda$4959 &  40.568 &  40.165 &  39.835 \\
{[\ionf{O}{iii}]}   $\lambda$5007 &  41.043 &  40.640 &  40.310 \\
\noalign{\medskip}
\multicolumn{4}{c}{                          \ldots }\\
\noalign{\medskip}
\noalign{\medskip}
\hline
\end{tabular}
\end{center}
\begin{list}{}{}
\footnotesize
\item[\textbf{Note:}]Table \ref{tab:cloudy:output} is published in its entirety in the machine-readable format. A portion is shown here for guidance regarding its form and content.
\item[]Model parameters for this example are as follows: metallicity $Z/$Z$_{\odot}=0.5$, mass-loading rate $\dot{M}_{\rm sc} = 0.607 \times 10^{-2}$ M$_{\odot}$\,yr$^{-1}$, actual wind velocity $V_{\infty}= 457$ km\,s$^{-1}$, SSC radius $R_{\rm sc} = 1$ pc,
stellar mass $M_{\star}= 2.05 \times 10^6$\,M$_{\odot}$, age $t=1$ Myr, ambient density $n_{\rm amb} = 100$ cm$^{-3}$, and ambient temperature $T_{\rm amb}$ estimated by \cloudy. 
\end{list}
\end{table}
 

\begin{itemize}
 \item[--] \verb|metal|: metallicity $\hat{Z} \equiv Z/$Z$_{\odot} = 1$, $0.5$, $0.25$, $0.125$.

 \item[--] \verb|dMdt|: mass-loading rate $\dot{M}_{\rm sc} = 10^{-1}$, $10^{-2}$, $10^{-3}$, $10^{-4} \times \hat{Z}^{0.72}$ M$_{\odot}$\,yr$^{-1}$.

 \item[--] \verb|Vinf|: wind terminal velocity $V_{\infty}= 250$, $500$, $1000 \times \hat{Z}^{0.13}$ km\,s$^{-1}$.

 \item[--] \verb|Rsc|: SSC radius $R_{\rm sc} = 1$ pc.

 \item[--] \verb|age|: current age $t = 1$ Myr.

 \item[--] \verb|Mstar|: total stellar mass $M_{\star}= 2.05 \times 10^6$\,M$_{\odot}$.

 \item[--] \verb|logLion|: ionizing luminosity $\log L_{\rm ion}$ (erg/s).
   
 \item[--] \verb|Namb|: ambient density $n_{\rm amb} = 1$, 10, $10^2$, $10^3$ cm$^{-3}$.

 \item[--] \verb|Tamb|: mean ambient temperature $T_{\rm amb}$ from \cloudy.

 \item[--] \verb|Rmax|: maximum radius $R_{\rm max}$ (pc) for the surface brightness integration.

 \item[--] \verb|Raper|: aperture radius $R_{\rm aper}$ (pc) for the luminosity integration.

 \item[--] \verb|Rshell|: shell exterior radius $R_{\rm shell}$ (pc).
 
 \item[--] \verb|Rstr|: Str\"{o}mgren radius $R_{\rm str}$ (pc) from \cloudy.
 
 \item[--] \verb|Rbin|: bubble interior radius $R_{\rm b,in}$ (pc).
 
 \item[--] \verb|Rbout|: bubble exterior radius $R_{\rm b,out}$ (pc) or shell interior radius.
 
 \item[--] \verb|Tbubble|: median temperature $T_{\rm bubble}$ of the bubble.

 \item[--] \verb|Tadi|: median adiabatic temperature $T_{\rm adi,med}$ of the expanding wind.
 
 \item[--] \verb|Twind|: median radiative temperature $T_{\rm w,med}$ of the expanding wind.
 
 \item[--] \verb|logUsp|: logarithmic ionization parameter $\log U_{\rm sph}$ in a spherical geometry from \cloudy.
 
 \item[--] \verb|thin|: optically thin (1) or thick (0) model.
 
 \item[--] \verb|mode|: the cooling/heating radiative/adiabatic modes: 
 1 (AW: adiabatic wind), 
 2 (AB: adiabatic bubble), 3 (AP: adiabatic, pressure-confined), 
 4 (CC: catastrophic cooling), 5 (CB: catastrophic cooling bubble), 
 and 6 (CP: catastrophic cooling, pressure-confined), described in details by \citet[][]{Danehkar2021}. 
 
  \item[--] \verb|H_1_1216|, \verb|H_1_6563|, \ldots , \verb|Ar_5_7006|: luminosities of the emission lines 
  Ly$\alpha$ $\lambda$1216 {\AA}, H$\alpha$ $\lambda$6563 {\AA}, \ldots , [\ionf{Ar}{v}] $\lambda$7006 {\AA},
  respectively. 
  
\end{itemize}

%

\vspace{1mm}


\software{\flash \citep{Fryxell2000}, yt \citep{Turk2011}, \cloudy \citep{Ferland2017}, Starburst99 \citep{Leitherer2014}, NumPy \citep{Harris2020}, SciPy \citep{Virtanen2020}, Matplotlib \citep{Hunter2007}.}

\end{appendix}

{ \small 
\begin{center}
\textbf{ORCID iDs}
\end{center}
\vspace{-5pt}

\noindent A.~Danehkar \orcidauthor{0000-0003-4552-5997} \url{https://orcid.org/0000-0003-4552-5997}

\noindent M. S. Oey \orcidauthor{0000-0002-5808-1320} \url{https://orcid.org/0000-0002-5808-1320}

\noindent W. J. Gray \orcidauthor{0000-0001-9014-3125} \url{https://orcid.org/0000-0001-9014-3125}

}



\begin{thebibliography}{108}
\expandafter\ifx\csname natexlab\endcsname\relax\def\natexlab#1{#1}\fi

\bibitem[{{Amor{\'\i}n} {et~al.}(2017){Amor{\'\i}n}, {Fontana},
  {P{\'e}rez-Montero}, {Castellano}, {Guaita}, {Grazian}, {Le F{\`e}vre},
  {Ribeiro}, {Schaerer}, {Tasca}, {Thomas}, {Bardelli}, {Cassar{\`a}},
  {Cassata}, {Cimatti}, {Contini}, {de Barros}, {Garilli}, {Giavalisco},
  {Hathi}, {Koekemoer}, {Le Brun}, {Lemaux}, {Maccagni}, {Pentericci}, {Pforr},
  {Talia}, {Tresse}, {Vanzella}, {Vergani}, {Zamorani}, {Zucca}, \&
  {Merlin}}]{Amorin2017}
{Amor{\'\i}n}, R., {Fontana}, A., {P{\'e}rez-Montero}, E., {et~al.} 2017,
  {\href{https://dx.doi.org/10.1038/s41550-017-0052}{\color{magenta}Nature
  Astron.}}, \href{https://ui.adsabs.harvard.edu/abs/2017NatAs...1E..52A}{1,
  0052}

\bibitem[{{Badnell}(2006)}]{Badnell2006}
{Badnell}, N.~R. 2006,
  {\href{https://dx.doi.org/10.1086/508465}{\color{magenta}\apjs}},
  \href{https://ui.adsabs.harvard.edu/abs/2006ApJS..167..334B}{167, 334}

\bibitem[{{Berg} {et~al.}(2019{\natexlab{a}}){Berg}, {Chisholm}, {Erb},
  {Pogge}, {Henry}, \& {Olivier}}]{Berg2019a}
{Berg}, D.~A., {Chisholm}, J., {Erb}, D.~K., {et~al.} 2019{\natexlab{a}},
  {\href{https://dx.doi.org/10.3847/2041-8213/ab21dc}{\color{magenta}\apjl}},
  \href{https://ui.adsabs.harvard.edu/abs/2019ApJ...878L...3B}{878, L3}

\bibitem[{{Berg} {et~al.}(2019{\natexlab{b}}){Berg}, {Erb}, {Henry},
  {Skillman}, \& {McQuinn}}]{Berg2019b}
{Berg}, D.~A., {Erb}, D.~K., {Henry}, R. B.~C., {Skillman}, E.~D., \&
  {McQuinn}, K. B.~W. 2019{\natexlab{b}},
  {\href{https://dx.doi.org/10.3847/1538-4357/ab020a}{\color{magenta}\apj}},
  \href{https://ui.adsabs.harvard.edu/abs/2019ApJ...874...93B}{874, 93}

\bibitem[{{Berg} {et~al.}(2016){Berg}, {Skillman}, {Henry}, {Erb}, \&
  {Carigi}}]{Berg2016}
{Berg}, D.~A., {Skillman}, E.~D., {Henry}, R. B.~C., {Erb}, D.~K., \& {Carigi},
  L. 2016,
  {\href{https://dx.doi.org/10.3847/0004-637X/827/2/126}{\color{magenta}\apj}},
  \href{https://ui.adsabs.harvard.edu/abs/2016ApJ...827..126B}{827, 126}

\bibitem[{{Bergvall} {et~al.}(2006){Bergvall}, {Zackrisson}, {Andersson},
  {Arnberg}, {Masegosa}, \& {{\"O}stlin}}]{Bergvall2006}
{Bergvall}, N., {Zackrisson}, E., {Andersson}, B.~G., {et~al.} 2006,
  {\href{https://dx.doi.org/10.1051/0004-6361:20053788}{\color{magenta}\aap}},
  \href{https://ui.adsabs.harvard.edu/abs/2006A&A...448..513B}{448, 513}

\bibitem[{{Bolatto} {et~al.}(2013){Bolatto}, {Warren}, {Leroy}, {Walter},
  {Veilleux}, {Ostriker}, {Ott}, {Zwaan}, {Fisher}, {Weiss}, {Rosolowsky}, \&
  {Hodge}}]{Bolatto2013}
{Bolatto}, A.~D., {Warren}, S.~R., {Leroy}, A.~K., {et~al.} 2013,
  {\href{https://dx.doi.org/10.1038/nature12351}{\color{magenta}\nat}},
  \href{https://ui.adsabs.harvard.edu/abs/2013Natur.499..450B}{499, 450}

\bibitem[{{Bryans} {et~al.}(2006){Bryans}, {Badnell}, {Gorczyca}, {Laming},
  {Mitthumsiri}, \& {Savin}}]{Bryans2006}
{Bryans}, P., {Badnell}, N.~R., {Gorczyca}, T.~W., {et~al.} 2006,
  {\href{https://dx.doi.org/10.1086/507629}{\color{magenta}\apjs}},
  \href{https://ui.adsabs.harvard.edu/abs/2006ApJS..167..343B}{167, 343}

\bibitem[{{Cant{\'o}} {et~al.}(2000){Cant{\'o}}, {Raga}, \&
  {Rodr{\'\i}guez}}]{Canto2000}
{Cant{\'o}}, J., {Raga}, A.~C., \& {Rodr{\'\i}guez}, L.~F. 2000,
  {\href{https://dx.doi.org/10.1086/308983}{\color{magenta}\apj}},
  \href{https://ui.adsabs.harvard.edu/abs/2000ApJ...536..896C}{536, 896}

\bibitem[{{Castor} {et~al.}(1975){Castor}, {McCray}, \& {Weaver}}]{Castor1975}
{Castor}, J., {McCray}, R., \& {Weaver}, R. 1975,
  {\href{https://dx.doi.org/10.1086/181908}{\color{magenta}\apjl}},
  \href{https://ui.adsabs.harvard.edu/abs/1975ApJ...200L.107C}{200, L107}

\bibitem[{{Chevalier} \& {Clegg}(1985)}]{Chevalier1985}
{Chevalier}, R.~A. \& {Clegg}, A.~W. 1985,
  {\href{https://dx.doi.org/10.1038/317044a0}{\color{magenta}\nat}},
  \href{https://ui.adsabs.harvard.edu/abs/1985Natur.317...44C}{317, 44}

\bibitem[{{Cottle} {et~al.}(2018){Cottle}, {Scannapieco}, \&
  {Br{\"u}ggen}}]{Cottle2018}
{Cottle}, J.~N., {Scannapieco}, E., \& {Br{\"u}ggen}, M. 2018,
  {\href{https://dx.doi.org/10.3847/1538-4357/aad55c}{\color{magenta}\apj}},
  \href{https://ui.adsabs.harvard.edu/abs/2018ApJ...864...96C}{864, 96}

\bibitem[{{Cox} \& {Tucker}(1969)}]{Cox1969}
{Cox}, D.~P. \& {Tucker}, W.~H. 1969,
  {\href{https://dx.doi.org/10.1086/150144}{\color{magenta}\apj}},
  \href{https://ui.adsabs.harvard.edu/abs/1969ApJ...157.1157C}{157, 1157}

\bibitem[{{Danehkar} {et~al.}(2021){Danehkar}, {Oey}, \& {Gray}}]{Danehkar2021}
{Danehkar}, A., {Oey}, M.~S., \& {Gray}, W.~J. 2021,
  {\href{https://dx.doi.org/10.3847/1538-4357/ac1a76}{\color{magenta}\apj}},
  \href{https://ui.adsabs.harvard.edu/abs/2021ApJ...921...91D}{921, 91}, (Paper
  II)

\bibitem[{{de Avillez} \& {Breitschwerdt}(2012)}]{deAvillez2012}
{de Avillez}, M.~A. \& {Breitschwerdt}, D. 2012,
  {\href{https://dx.doi.org/10.1088/2041-8205/761/2/L19}{\color{magenta}\apjl}},
  \href{https://ui.adsabs.harvard.edu/abs/2012ApJ...761L..19D}{761, L19}

\bibitem[{{De Vis} {et~al.}(2019){De Vis}, {Jones}, {Viaene}, {Casasola},
  {Clark}, {Baes}, {Bianchi}, {Cassara}, {Davies}, {De Looze}, {Galametz},
  {Galliano}, {Lianou}, {Madden}, {Manilla-Robles}, {Mosenkov}, {Nersesian},
  {Roychowdhury}, {Xilouris}, \& {Ysard}}]{DeVis2019}
{De Vis}, P., {Jones}, A., {Viaene}, S., {et~al.} 2019,
  {\href{https://dx.doi.org/10.1051/0004-6361/201834444}{\color{magenta}\aap}},
  \href{https://ui.adsabs.harvard.edu/abs/2019A&A...623A...5D}{623, A5}

\bibitem[{{della Ceca} {et~al.}(1996){della Ceca}, {Griffiths}, {Heckman}, \&
  {MacKenty}}]{dellaCeca1996}
{della Ceca}, R., {Griffiths}, R.~E., {Heckman}, T.~M., \& {MacKenty}, J.~W.
  1996, {\href{https://dx.doi.org/10.1086/177813}{\color{magenta}\apj}},
  \href{https://ui.adsabs.harvard.edu/abs/1996ApJ...469..662D}{469, 662}

\bibitem[{{Dopita} \& {Sutherland}(2003)}]{Dopita2003}
{Dopita}, M.~A. \& {Sutherland}, R.~S. 2003, {Astrophysics of the Diffuse
  Universe} (Berlin, New York: Springer)

\bibitem[{{Ekstr{\"o}m} {et~al.}(2012){Ekstr{\"o}m}, {Georgy}, {Eggenberger},
  {Meynet}, {Mowlavi}, {Wyttenbach}, {Granada}, {Decressin}, {Hirschi},
  {Frischknecht}, {Charbonnel}, \& {Maeder}}]{Ekstroem2012}
{Ekstr{\"o}m}, S., {Georgy}, C., {Eggenberger}, P., {et~al.} 2012,
  {\href{https://dx.doi.org/10.1051/0004-6361/201117751}{\color{magenta}\aap}},
  \href{https://ui.adsabs.harvard.edu/abs/2012A&A...537A.146E}{537, A146}

\bibitem[{{Fabian} {et~al.}(1984){Fabian}, {Nulsen}, \&
  {Canizares}}]{Fabian1984}
{Fabian}, A.~C., {Nulsen}, P.~E.~J., \& {Canizares}, C.~R. 1984,
  {\href{https://dx.doi.org/10.1038/310733a0}{\color{magenta}\nat}},
  \href{https://ui.adsabs.harvard.edu/abs/1984Natur.310..733F}{310, 733}

\bibitem[{{Feltre} {et~al.}(2016){Feltre}, {Charlot}, \& {Gutkin}}]{Feltre2016}
{Feltre}, A., {Charlot}, S., \& {Gutkin}, J. 2016,
  {\href{https://dx.doi.org/10.1093/mnras/stv2794}{\color{magenta}\mnras}},
  \href{https://ui.adsabs.harvard.edu/abs/2016MNRAS.456.3354F}{456, 3354}

\bibitem[{{Ferland} {et~al.}(2017){Ferland}, {Chatzikos}, {Guzm{\'a}n},
  {Lykins}, {van Hoof}, {Williams}, {Abel}, {Badnell}, {Keenan}, {Porter}, \&
  {Stancil}}]{Ferland2017}
{Ferland}, G.~J., {Chatzikos}, M., {Guzm{\'a}n}, F., {et~al.} 2017, \rmxaa,
  \href{https://ui.adsabs.harvard.edu/abs/2017RMxAA..53..385F}{53, 385}

\bibitem[{{Ferland} {et~al.}(1998){Ferland}, {Korista}, {Verner}, {Ferguson},
  {Kingdon}, \& {Verner}}]{Ferland1998}
{Ferland}, G.~J., {Korista}, K.~T., {Verner}, D.~A., {et~al.} 1998,
  {\href{https://dx.doi.org/10.1086/316190}{\color{magenta}\pasp}},
  \href{https://ui.adsabs.harvard.edu/abs/1998PASP..110..761F}{110, 761}

\bibitem[{{Ferland} {et~al.}(2013){Ferland}, {Porter}, {van Hoof}, {Williams},
  {Abel}, {Lykins}, {Shaw}, {Henney}, \& {Stancil}}]{Ferland2013}
{Ferland}, G.~J., {Porter}, R.~L., {van Hoof}, P.~A.~M., {et~al.} 2013, \rmxaa,
  \href{https://ui.adsabs.harvard.edu/abs/2013RMxAA..49..137F}{49, 137}

\bibitem[{{Fryxell} {et~al.}(2000){Fryxell}, {Olson}, {Ricker}, {Timmes},
  {Zingale}, {Lamb}, {MacNeice}, {Rosner}, {Truran}, \& {Tufo}}]{Fryxell2000}
{Fryxell}, B., {Olson}, K., {Ricker}, P., {et~al.} 2000,
  {\href{https://dx.doi.org/10.1086/317361}{\color{magenta}\apjs}},
  \href{https://ui.adsabs.harvard.edu/abs/2000ApJS..131..273F}{131, 273}

\bibitem[{{Garnett} {et~al.}(1999){Garnett}, {Shields}, {Peimbert},
  {Torres-Peimbert}, {Skillman}, {Dufour}, {Terlevich}, \&
  {Terlevich}}]{Garnett1999}
{Garnett}, D.~R., {Shields}, G.~A., {Peimbert}, M., {et~al.} 1999,
  {\href{https://dx.doi.org/10.1086/306860}{\color{magenta}\apj}},
  \href{https://ui.adsabs.harvard.edu/abs/1999ApJ...513..168G}{513, 168}

\bibitem[{{Georgy} {et~al.}(2013){Georgy}, {Ekstr{\"o}m}, {Eggenberger},
  {Meynet}, {Haemmerl{\'e}}, {Maeder}, {Granada}, {Groh}, {Hirschi}, {Mowlavi},
  {Yusof}, {Charbonnel}, {Decressin}, \& {Barblan}}]{Georgy2013}
{Georgy}, C., {Ekstr{\"o}m}, S., {Eggenberger}, P., {et~al.} 2013,
  {\href{https://dx.doi.org/10.1051/0004-6361/201322178}{\color{magenta}\aap}},
  \href{https://ui.adsabs.harvard.edu/abs/2013A&A...558A.103G}{558, A103}

\bibitem[{{Georgy} {et~al.}(2012){Georgy}, {Ekstr{\"o}m}, {Meynet}, {Massey},
  {Levesque}, {Hirschi}, {Eggenberger}, \& {Maeder}}]{Georgy2012}
{Georgy}, C., {Ekstr{\"o}m}, S., {Meynet}, G., {et~al.} 2012,
  {\href{https://dx.doi.org/10.1051/0004-6361/201118340}{\color{magenta}\aap}},
  \href{https://ui.adsabs.harvard.edu/abs/2012A&A...542A..29G}{542, A29}

\bibitem[{{Gnat} \& {Ferland}(2012)}]{Gnat2012}
{Gnat}, O. \& {Ferland}, G.~J. 2012,
  {\href{https://dx.doi.org/10.1088/0067-0049/199/1/20}{\color{magenta}\apjs}},
  \href{https://ui.adsabs.harvard.edu/abs/2012ApJS..199...20G}{199, 20}

\bibitem[{{Gnat} \& {Sternberg}(2007)}]{Gnat2007}
{Gnat}, O. \& {Sternberg}, A. 2007,
  {\href{https://dx.doi.org/10.1086/509786}{\color{magenta}\apjs}},
  \href{https://ui.adsabs.harvard.edu/abs/2007ApJS..168..213G}{168, 213}

\bibitem[{{Gray} {et~al.}(2019{\natexlab{a}}){Gray}, {Oey}, {Silich}, \&
  {Scannapieco}}]{Gray2019a}
{Gray}, W.~J., {Oey}, M.~S., {Silich}, S., \& {Scannapieco}, E.
  2019{\natexlab{a}},
  {\href{https://dx.doi.org/10.3847/1538-4357/ab510d}{\color{magenta}\apj}},
  \href{https://ui.adsabs.harvard.edu/abs/2019ApJ...887..161G}{887, 161},
  (Paper I)

\bibitem[{{Gray} \& {Scannapieco}(2016)}]{Gray2016}
{Gray}, W.~J. \& {Scannapieco}, E. 2016,
  {\href{https://dx.doi.org/10.3847/0004-637X/818/2/198}{\color{magenta}\apj}},
  \href{https://ui.adsabs.harvard.edu/abs/2016ApJ...818..198G}{818, 198}

\bibitem[{{Gray} {et~al.}(2015){Gray}, {Scannapieco}, \& {Kasen}}]{Gray2015}
{Gray}, W.~J., {Scannapieco}, E., \& {Kasen}, D. 2015,
  {\href{https://dx.doi.org/10.1088/0004-637X/801/2/107}{\color{magenta}\apj}},
  \href{https://ui.adsabs.harvard.edu/abs/2015ApJ...801..107G}{801, 107}

\bibitem[{{Gray} {et~al.}(2019{\natexlab{b}}){Gray}, {Scannapieco}, \&
  {Lehnert}}]{Gray2019}
{Gray}, W.~J., {Scannapieco}, E., \& {Lehnert}, M.~D. 2019{\natexlab{b}},
  {\href{https://dx.doi.org/10.3847/1538-4357/ab1004}{\color{magenta}\apj}},
  \href{https://ui.adsabs.harvard.edu/abs/2019ApJ...875..110G}{875, 110}

\bibitem[{{Grimes} {et~al.}(2007){Grimes}, {Heckman}, {Strickland}, {Dixon},
  {Sembach}, {Overzier}, {Hoopes}, {Aloisi}, \& {Ptak}}]{Grimes2007}
{Grimes}, J.~P., {Heckman}, T., {Strickland}, D., {et~al.} 2007,
  {\href{https://dx.doi.org/10.1086/521353}{\color{magenta}\apj}},
  \href{https://ui.adsabs.harvard.edu/abs/2007ApJ...668..891G}{668, 891}

\bibitem[{{Gutkin} {et~al.}(2016){Gutkin}, {Charlot}, \&
  {Bruzual}}]{Gutkin2016}
{Gutkin}, J., {Charlot}, S., \& {Bruzual}, G. 2016,
  {\href{https://dx.doi.org/10.1093/mnras/stw1716}{\color{magenta}\mnras}},
  \href{https://ui.adsabs.harvard.edu/abs/2016MNRAS.462.1757G}{462, 1757}

\bibitem[{{Harris} {et~al.}(2020){Harris}, {Millman}, {van der Walt},
  {Gommers}, {Virtanen}, {Cournapeau}, {Wieser}, {Taylor}, {Berg}, {Smith},
  {Kern}, {Picus}, {Hoyer}, {van Kerkwijk}, {Brett}, {Haldane}, {del R{\'\i}o},
  {Wiebe}, {Peterson}, {G{\'e}rard-Marchant}, {Sheppard}, {Reddy}, {Weckesser},
  {Abbasi}, {Gohlke}, \& {Oliphant}}]{Harris2020}
{Harris}, C.~R., {Millman}, K.~J., {van der Walt}, S.~J., {et~al.} 2020,
  {\href{https://dx.doi.org/10.1038/s41586-020-2649-2}{\color{magenta}\nat}},
  \href{https://ui.adsabs.harvard.edu/abs/2020Natur.585..357H}{585, 357}

\bibitem[{{Hayes} {et~al.}(2016){Hayes}, {Melinder}, {{\"O}stlin}, {Scarlata},
  {Lehnert}, \& {Mannerstr{\"o}m-Jansson}}]{Hayes2016}
{Hayes}, M., {Melinder}, J., {{\"O}stlin}, G., {et~al.} 2016,
  {\href{https://dx.doi.org/10.3847/0004-637X/828/1/49}{\color{magenta}\apj}},
  \href{https://ui.adsabs.harvard.edu/abs/2016ApJ...828...49H}{828, 49}

\bibitem[{{Heckman} {et~al.}(1990){Heckman}, {Armus}, \& {Miley}}]{Heckman1990}
{Heckman}, T.~M., {Armus}, L., \& {Miley}, G.~K. 1990,
  {\href{https://dx.doi.org/10.1086/191522}{\color{magenta}\apjs}},
  \href{https://ui.adsabs.harvard.edu/abs/1990ApJS...74..833H}{74, 833}

\bibitem[{{Heckman} {et~al.}(2001){Heckman}, {Sembach}, {Meurer}, {Strickland
  }, {Martin}, {Calzetti}, \& {Leitherer}}]{Heckman2001}
{Heckman}, T.~M., {Sembach}, K.~R., {Meurer}, G.~R., {et~al.} 2001,
  {\href{https://dx.doi.org/10.1086/321422}{\color{magenta}\apj}},
  \href{https://ui.adsabs.harvard.edu/abs/2001ApJ...554.1021H}{554, 1021}

\bibitem[{{Hillier} \& {Miller}(1998)}]{Hillier1998}
{Hillier}, D.~J. \& {Miller}, D.~L. 1998,
  {\href{https://dx.doi.org/10.1086/305350}{\color{magenta}\apj}},
  \href{https://ui.adsabs.harvard.edu/abs/1998ApJ...496..407H}{496, 407}

\bibitem[{{Hirschmann} {et~al.}(2019){Hirschmann}, {Charlot}, {Feltre}, {Naab},
  {Somerville}, \& {Choi}}]{Hirschmann2019}
{Hirschmann}, M., {Charlot}, S., {Feltre}, A., {et~al.} 2019,
  {\href{https://dx.doi.org/10.1093/mnras/stz1256}{\color{magenta}\mnras}},
  \href{https://ui.adsabs.harvard.edu/abs/2019MNRAS.487..333H}{487, 333}

\bibitem[{{Hunter}(2007)}]{Hunter2007}
{Hunter}, J.~D. 2007,
  {\href{https://dx.doi.org/10.1109/MCSE.2007.55}{\color{magenta}Comput. Sci.
  Eng.}}, \href{https://ui.adsabs.harvard.edu/abs/2007CSE.....9...90H}{9, 90}

\bibitem[{{Izotov} \& {Thuan}(1999)}]{Izotov1999}
{Izotov}, Y.~I. \& {Thuan}, T.~X. 1999,
  {\href{https://dx.doi.org/10.1086/306708}{\color{magenta}\apj}},
  \href{https://ui.adsabs.harvard.edu/abs/1999ApJ...511..639I}{511, 639}

\bibitem[{{Izotov} {et~al.}(2018){Izotov}, {Worseck}, {Schaerer}, {Guseva},
  {Thuan}, {Fricke}, \& {Orlitov{\'a}}}]{Izotov2018a}
{Izotov}, Y.~I., {Worseck}, G., {Schaerer}, D., {et~al.} 2018,
  {\href{https://dx.doi.org/10.1093/mnras/sty1378}{\color{magenta}\mnras}},
  \href{https://ui.adsabs.harvard.edu/abs/2018MNRAS.478.4851I}{478, 4851}

\bibitem[{{James} {et~al.}(2009){James}, {Tsamis}, {Barlow}, {Westmoquette},
  {Walsh}, {Cuisinier}, \& {Exter}}]{James2009}
{James}, B.~L., {Tsamis}, Y.~G., {Barlow}, M.~J., {et~al.} 2009,
  {\href{https://dx.doi.org/10.1111/j.1365-2966.2009.15172.x}{\color{magenta}\mnras}},
  \href{https://ui.adsabs.harvard.edu/abs/2009MNRAS.398....2J}{398, 2}

\bibitem[{{James} {et~al.}(2013){James}, {Tsamis}, {Walsh}, {Barlow}, \&
  {Westmoquette}}]{James2013}
{James}, B.~L., {Tsamis}, Y.~G., {Walsh}, J.~R., {Barlow}, M.~J., \&
  {Westmoquette}, M.~S. 2013,
  {\href{https://dx.doi.org/10.1093/mnras/stt034}{\color{magenta}\mnras}},
  \href{https://ui.adsabs.harvard.edu/abs/2013MNRAS.430.2097J}{430, 2097}

\bibitem[{{Jaskot} {et~al.}(2017){Jaskot}, {Oey}, {Scarlata}, \&
  {Dowd}}]{Jaskot2017}
{Jaskot}, A.~E., {Oey}, M.~S., {Scarlata}, C., \& {Dowd}, T. 2017,
  {\href{https://dx.doi.org/10.3847/2041-8213/aa9d83}{\color{magenta}\apjl}},
  \href{https://ui.adsabs.harvard.edu/abs/2017ApJ...851L...9J}{851, L9}

\bibitem[{{Kafatos}(1973)}]{Kafatos1973}
{Kafatos}, M. 1973,
  {\href{https://dx.doi.org/10.1086/152151}{\color{magenta}\apj}},
  \href{https://ui.adsabs.harvard.edu/abs/1973ApJ...182..433K}{182, 433}

\bibitem[{{Krause} {et~al.}(2016){Krause}, {Charbonnel}, {Bastian}, \&
  {Diehl}}]{Krause2016}
{Krause}, M. G.~H., {Charbonnel}, C., {Bastian}, N., \& {Diehl}, R. 2016,
  {\href{https://dx.doi.org/10.1051/0004-6361/201526685}{\color{magenta}\aap}},
  \href{https://ui.adsabs.harvard.edu/abs/2016A&A...587A..53K}{587, A53}

\bibitem[{{Lee}(2013)}]{Lee2013}
{Lee}, D. 2013,
  {\href{https://dx.doi.org/10.1016/j.jcp.2013.02.049}{\color{magenta}J.
  Comput. Phys.}},
  \href{https://ui.adsabs.harvard.edu/abs/2013JCoPh.243..269L}{243, 269}

\bibitem[{{Lee} \& {Deane}(2009)}]{Lee2009a}
{Lee}, D. \& {Deane}, A.~E. 2009,
  {\href{https://dx.doi.org/10.1016/j.jcp.2008.08.026}{\color{magenta}J.
  Comput. Phys.}},
  \href{https://ui.adsabs.harvard.edu/abs/2009JCoPh.228..952L}{228, 952}

\bibitem[{{Lee} {et~al.}(2009){Lee}, {Deane}, \& {Federrath}}]{Lee2009}
{Lee}, D., {Deane}, A.~E., \& {Federrath}, C. 2009, ASP Conf. Ser., Vol. 406,
  {A New Multidimensional Unsplit MHD Solver in FLASH3}, ed. N.~V. {Pogorelov},
  E.~{Audit}, P.~{Colella}, \& G.~P. {Zank}, 243

\bibitem[{{Leitherer} {et~al.}(2014){Leitherer}, {Ekstr{\"o}m}, {Meynet},
  {Schaerer}, {Agienko}, \& {Levesque}}]{Leitherer2014}
{Leitherer}, C., {Ekstr{\"o}m}, S., {Meynet}, G., {et~al.} 2014,
  {\href{https://dx.doi.org/10.1088/0067-0049/212/1/14}{\color{magenta}\apjs}},
  \href{https://ui.adsabs.harvard.edu/abs/2014ApJS..212...14L}{212, 14}

\bibitem[{{Leitherer} {et~al.}(1999){Leitherer}, {Schaerer}, {Goldader},
  {Delgado}, {Robert}, {Kune}, {de Mello}, {Devost}, \&
  {Heckman}}]{Leitherer1999}
{Leitherer}, C., {Schaerer}, D., {Goldader}, J.~D., {et~al.} 1999,
  {\href{https://dx.doi.org/10.1086/313233}{\color{magenta}\apjs}},
  \href{https://ui.adsabs.harvard.edu/abs/1999ApJS..123....3L}{123, 3}

\bibitem[{{Leroy} {et~al.}(2015){Leroy}, {Walter}, {Martini}, {Roussel},
  {Sandstrom}, {Ott}, {Weiss}, {Bolatto}, {Schuster}, \&
  {Dessauges-Zavadsky}}]{Leroy2015}
{Leroy}, A.~K., {Walter}, F., {Martini}, P., {et~al.} 2015,
  {\href{https://dx.doi.org/10.1088/0004-637X/814/2/83}{\color{magenta}\apj}},
  \href{https://ui.adsabs.harvard.edu/abs/2015ApJ...814...83L}{814, 83}

\bibitem[{{Levesque} {et~al.}(2012){Levesque}, {Leitherer}, {Ekstrom},
  {Meynet}, \& {Schaerer}}]{Levesque2012}
{Levesque}, E.~M., {Leitherer}, C., {Ekstrom}, S., {Meynet}, G., \& {Schaerer},
  D. 2012,
  {\href{https://dx.doi.org/10.1088/0004-637X/751/1/67}{\color{magenta}\apj}},
  \href{https://ui.adsabs.harvard.edu/abs/2012ApJ...751...67L}{751, 67}

\bibitem[{{Li} {et~al.}(2017){Li}, {Bryan}, \& {Ostriker}}]{Li2017}
{Li}, M., {Bryan}, G.~L., \& {Ostriker}, J.~P. 2017,
  {\href{https://dx.doi.org/10.3847/2041-8213/835/1/L10}{\color{magenta}\apjl}},
  \href{https://ui.adsabs.harvard.edu/abs/2017ApJ...835L..10L}{835, L10}

\bibitem[{{Lochhaas} {et~al.}(2020){Lochhaas}, {Bryan}, {Li}, {Li}, \&
  {Fielding}}]{Lochhaas2020}
{Lochhaas}, C., {Bryan}, G.~L., {Li}, Y., {Li}, M., \& {Fielding}, D. 2020,
  {\href{https://dx.doi.org/10.1093/mnras/staa358}{\color{magenta}\mnras}},
  \href{https://ui.adsabs.harvard.edu/abs/2020MNRAS.493.1461L}{493, 1461}

\bibitem[{{Lochhaas} {et~al.}(2018){Lochhaas}, {Thompson}, {Quataert}, \&
  {Weinberg}}]{Lochhaas2018}
{Lochhaas}, C., {Thompson}, T.~A., {Quataert}, E., \& {Weinberg}, D.~H. 2018,
  {\href{https://dx.doi.org/10.1093/mnras/sty2421}{\color{magenta}\mnras}},
  \href{https://ui.adsabs.harvard.edu/abs/2018MNRAS.481.1873L}{481, 1873}

\bibitem[{{Marques-Chaves} {et~al.}(2021){Marques-Chaves}, {Schaerer},
  {{\'A}lvarez-M{\'a}rquez}, {Colina}, {Dessauges-Zavadsky},
  {P{\'e}rez-Fournon}, {Saldana-Lopez}, \& {Verhamme}}]{Marques-Chaves2021}
{Marques-Chaves}, R., {Schaerer}, D., {{\'A}lvarez-M{\'a}rquez}, J., {et~al.}
  2021,
  {\href{https://dx.doi.org/10.1093/mnras/stab2187}{\color{magenta}\mnras}},
  \href{https://ui.adsabs.harvard.edu/abs/2021MNRAS.507..524M}{507, 524}

\bibitem[{{Martin}(1999)}]{Martin1999}
{Martin}, C.~L. 1999,
  {\href{https://dx.doi.org/10.1086/306863}{\color{magenta}\apj}},
  \href{https://ui.adsabs.harvard.edu/abs/1999ApJ...513..156M}{513, 156}

\bibitem[{{Masters} {et~al.}(2014){Masters}, {McCarthy}, {Siana}, {Malkan},
  {Mobasher}, {Atek}, {Henry}, {Martin}, {Rafelski}, {Hathi}, {Scarlata},
  {Ross}, {Bunker}, {Blanc}, {Bedregal}, {Dom{\'\i}nguez}, {Colbert},
  {Teplitz}, \& {Dressler}}]{Masters2014}
{Masters}, D., {McCarthy}, P., {Siana}, B., {et~al.} 2014,
  {\href{https://dx.doi.org/10.1088/0004-637X/785/2/153}{\color{magenta}\apj}},
  \href{https://ui.adsabs.harvard.edu/abs/2014ApJ...785..153M}{785, 153}

\bibitem[{{Meyer} {et~al.}(1998){Meyer}, {Jura}, \& {Cardelli}}]{Meyer1998}
{Meyer}, D.~M., {Jura}, M., \& {Cardelli}, J.~A. 1998,
  {\href{https://dx.doi.org/10.1086/305128}{\color{magenta}\apj}},
  \href{https://ui.adsabs.harvard.edu/abs/1998ApJ...493..222M}{493, 222}

\bibitem[{{Mokiem} {et~al.}(2007){Mokiem}, {de Koter}, {Vink}, {Puls}, {Evans},
  {Smartt}, {Crowther}, {Herrero}, {Langer}, {Lennon}, {Najarro}, \&
  {Villamariz}}]{Mokiem2007}
{Mokiem}, M.~R., {de Koter}, A., {Vink}, J.~S., {et~al.} 2007,
  {\href{https://dx.doi.org/10.1051/0004-6361:20077545}{\color{magenta}\aap}},
  \href{https://ui.adsabs.harvard.edu/abs/2007A&A...473..603M}{473, 603}

\bibitem[{{Oey} {et~al.}(2017){Oey}, {Herrera}, {Silich}, {Reiter}, {James},
  {Jaskot}, \& {Micheva}}]{Oey2017}
{Oey}, M.~S., {Herrera}, C.~N., {Silich}, S., {et~al.} 2017,
  {\href{https://dx.doi.org/10.3847/2041-8213/aa9215}{\color{magenta}\apjl}},
  \href{https://ui.adsabs.harvard.edu/abs/2017ApJ...849L...1O}{849, L1}

\bibitem[{{Oppenheimer} \& {Schaye}(2013)}]{Oppenheimer2013}
{Oppenheimer}, B.~D. \& {Schaye}, J. 2013,
  {\href{https://dx.doi.org/10.1093/mnras/stt1043}{\color{magenta}\mnras}},
  \href{https://ui.adsabs.harvard.edu/abs/2013MNRAS.434.1043O}{434, 1043}

\bibitem[{{Ott} {et~al.}(2005{\natexlab{a}}){Ott}, {Walter}, \&
  {Brinks}}]{Ott2005a}
{Ott}, J., {Walter}, F., \& {Brinks}, E. 2005{\natexlab{a}},
  {\href{https://dx.doi.org/10.1111/j.1365-2966.2005.08863.x}{\color{magenta}\mnras}},
  \href{https://ui.adsabs.harvard.edu/abs/2005MNRAS.358.1453O}{358, 1453}

\bibitem[{{Ott} {et~al.}(2005{\natexlab{b}}){Ott}, {Weiss}, {Henkel}, \&
  {Walter}}]{Ott2005}
{Ott}, J., {Weiss}, A., {Henkel}, C., \& {Walter}, F. 2005{\natexlab{b}},
  {\href{https://dx.doi.org/10.1086/431661}{\color{magenta}\apj}},
  \href{https://ui.adsabs.harvard.edu/abs/2005ApJ...629..767O}{629, 767}

\bibitem[{{Otte} {et~al.}(2003){Otte}, {Murphy}, {Howk}, {Wang}, {Oegerle}, \&
  {Sembach}}]{Otte2003}
{Otte}, B., {Murphy}, E.~M., {Howk}, J.~C., {et~al.} 2003,
  {\href{https://dx.doi.org/10.1086/375535}{\color{magenta}\apj}},
  \href{https://ui.adsabs.harvard.edu/abs/2003ApJ...591..821O}{591, 821}

\bibitem[{{Pauldrach} {et~al.}(2001){Pauldrach}, {Hoffmann}, \&
  {Lennon}}]{Pauldrach2001}
{Pauldrach}, A.~W.~A., {Hoffmann}, T.~L., \& {Lennon}, M. 2001,
  {\href{https://dx.doi.org/10.1051/0004-6361:20010805}{\color{magenta}\aap}},
  \href{https://ui.adsabs.harvard.edu/abs/2001A&A...375..161P}{375, 161}

\bibitem[{{Raymond} {et~al.}(1976){Raymond}, {Cox}, \& {Smith}}]{Raymond1976}
{Raymond}, J.~C., {Cox}, D.~P., \& {Smith}, B.~W. 1976,
  {\href{https://dx.doi.org/10.1086/154170}{\color{magenta}\apj}},
  \href{https://ui.adsabs.harvard.edu/abs/1976ApJ...204..290R}{204, 290}

\bibitem[{{Richard} {et~al.}(2011){Richard}, {Jones}, {Ellis}, {Stark},
  {Livermore}, \& {Swinbank}}]{Richard2011}
{Richard}, J., {Jones}, T., {Ellis}, R., {et~al.} 2011,
  {\href{https://dx.doi.org/10.1111/j.1365-2966.2010.18161.x}{\color{magenta}\mnras}},
  \href{https://ui.adsabs.harvard.edu/abs/2011MNRAS.413..643R}{413, 643}

\bibitem[{{Richings} {et~al.}(2014){Richings}, {Schaye}, \&
  {Oppenheimer}}]{Richings2014}
{Richings}, A.~J., {Schaye}, J., \& {Oppenheimer}, B.~D. 2014,
  {\href{https://dx.doi.org/10.1093/mnras/stu525}{\color{magenta}\mnras}},
  \href{https://ui.adsabs.harvard.edu/abs/2014MNRAS.440.3349R}{440, 3349}

\bibitem[{{Sarazin}(1988)}]{Sarazin1988}
{Sarazin}, C.~L. 1988, {X-ray emission from clusters of galaxies} (Cambridge:
  Cambridge University Press)

\bibitem[{{Savage} {et~al.}(1977){Savage}, {Bohlin}, {Drake}, \&
  {Budich}}]{Savage1977}
{Savage}, B.~D., {Bohlin}, R.~C., {Drake}, J.~F., \& {Budich}, W. 1977,
  {\href{https://dx.doi.org/10.1086/155471}{\color{magenta}\apj}},
  \href{https://ui.adsabs.harvard.edu/abs/1977ApJ...216..291S}{216, 291}

\bibitem[{{Schmutzler} \& {Tscharnuter}(1993)}]{Schmutzler1993}
{Schmutzler}, T. \& {Tscharnuter}, W.~M. 1993, \aap,
  \href{https://ui.adsabs.harvard.edu/abs/1993A&A...273..318S}{273, 318}

\bibitem[{{Schneider} {et~al.}(2018){Schneider}, {Robertson}, \&
  {Thompson}}]{Schneider2018}
{Schneider}, E.~E., {Robertson}, B.~E., \& {Thompson}, T.~A. 2018,
  {\href{https://dx.doi.org/10.3847/1538-4357/aacce1}{\color{magenta}\apj}},
  \href{https://ui.adsabs.harvard.edu/abs/2018ApJ...862...56S}{862, 56}

\bibitem[{{Senchyna} {et~al.}(2017){Senchyna}, {Stark}, {Vidal-Garc{\'\i}a},
  {Chevallard}, {Charlot}, {Mainali}, {Jones}, {Wofford}, {Feltre}, \&
  {Gutkin}}]{Senchyna2017}
{Senchyna}, P., {Stark}, D.~P., {Vidal-Garc{\'\i}a}, A., {et~al.} 2017,
  {\href{https://dx.doi.org/10.1093/mnras/stx2059}{\color{magenta}\mnras}},
  \href{https://ui.adsabs.harvard.edu/abs/2017MNRAS.472.2608S}{472, 2608}

\bibitem[{{Shapiro} \& {Moore}(1976)}]{Shapiro1976}
{Shapiro}, P.~R. \& {Moore}, R.~T. 1976,
  {\href{https://dx.doi.org/10.1086/154515}{\color{magenta}\apj}},
  \href{https://ui.adsabs.harvard.edu/abs/1976ApJ...207..460S}{207, 460}

\bibitem[{{Shull} \& {van Steenberg}(1982)}]{Shull1982}
{Shull}, J.~M. \& {van Steenberg}, M. 1982,
  {\href{https://dx.doi.org/10.1086/190769}{\color{magenta}\apjs}},
  \href{https://ui.adsabs.harvard.edu/abs/1982ApJS...48...95S}{48, 95}

\bibitem[{{Silich} \& {Tenorio-Tagle}(2017)}]{Silich2017}
{Silich}, S. \& {Tenorio-Tagle}, G. 2017,
  {\href{https://dx.doi.org/10.1093/mnras/stw2879}{\color{magenta}\mnras}},
  \href{https://ui.adsabs.harvard.edu/abs/2017MNRAS.465.1375S}{465, 1375}

\bibitem[{{Silich} {et~al.}(2003){Silich}, {Tenorio-Tagle}, \&
  {Mu{\~n}oz-Tu{\~n}{\'o}n}}]{Silich2003}
{Silich}, S., {Tenorio-Tagle}, G., \& {Mu{\~n}oz-Tu{\~n}{\'o}n}, C. 2003,
  {\href{https://dx.doi.org/10.1086/375133}{\color{magenta}\apj}},
  \href{https://ui.adsabs.harvard.edu/abs/2003ApJ...590..791S}{590, 791}

\bibitem[{{Silich} {et~al.}(2004){Silich}, {Tenorio-Tagle}, \&
  {Rodr{\'\i}guez-Gonz{\'a}lez}}]{Silich2004}
{Silich}, S., {Tenorio-Tagle}, G., \& {Rodr{\'\i}guez-Gonz{\'a}lez}, A. 2004,
  {\href{https://dx.doi.org/10.1086/421702}{\color{magenta}\apj}},
  \href{https://ui.adsabs.harvard.edu/abs/2004ApJ...610..226S}{610, 226}

\bibitem[{{Smith} {et~al.}(2006){Smith}, {Westmoquette}, {Gallagher},
  {O'Connell}, {Rosario}, \& {de Grijs}}]{Smith2006}
{Smith}, L.~J., {Westmoquette}, M.~S., {Gallagher}, J.~S., {et~al.} 2006,
  {\href{https://dx.doi.org/10.1111/j.1365-2966.2006.10507.x}{\color{magenta}\mnras}},
  \href{https://ui.adsabs.harvard.edu/abs/2006MNRAS.370..513S}{370, 513}

\bibitem[{{Strickland} {et~al.}(1997){Strickland}, {Ponman}, \&
  {Stevens}}]{Strickland1997}
{Strickland}, D.~K., {Ponman}, T.~J., \& {Stevens}, I.~R. 1997, \aap,
  \href{https://ui.adsabs.harvard.edu/abs/1997A&A...320..378S}{320, 378}

\bibitem[{{Sutherland} \& {Dopita}(1993)}]{Sutherland1993}
{Sutherland}, R.~S. \& {Dopita}, M.~A. 1993,
  {\href{https://dx.doi.org/10.1086/191823}{\color{magenta}\apjs}},
  \href{https://ui.adsabs.harvard.edu/abs/1993ApJS...88..253S}{88, 253}

\bibitem[{{Tenorio-Tagle} {et~al.}(2005){Tenorio-Tagle}, {Silich},
  {Rodr{\'\i}guez-Gonz{\'a}lez}, \&
  {Mu{\~n}oz-Tu{\~n}{\'o}n}}]{Tenorio-Tagle2005}
{Tenorio-Tagle}, G., {Silich}, S., {Rodr{\'\i}guez-Gonz{\'a}lez}, A., \&
  {Mu{\~n}oz-Tu{\~n}{\'o}n}, C. 2005,
  {\href{https://dx.doi.org/10.1086/426962}{\color{magenta}\apj}},
  \href{https://ui.adsabs.harvard.edu/abs/2005ApJ...620..217T}{620, 217}

\bibitem[{{Tenorio-Tagle} {et~al.}(2007){Tenorio-Tagle}, {W{\"u}nsch},
  {Silich}, \& {Palou{\v{s}}}}]{Tenorio-Tagle2007}
{Tenorio-Tagle}, G., {W{\"u}nsch}, R., {Silich}, S., \& {Palou{\v{s}}}, J.
  2007, {\href{https://dx.doi.org/10.1086/511671}{\color{magenta}\apj}},
  \href{https://ui.adsabs.harvard.edu/abs/2007ApJ...658.1196T}{658, 1196}

\bibitem[{{Toro} {et~al.}(1994){Toro}, {Spruce}, \& {Speares}}]{Toro1994}
{Toro}, E.~F., {Spruce}, M., \& {Speares}, W. 1994,
  {\href{https://dx.doi.org/10.1007/BF01414629}{\color{magenta}Shock Waves}},
  \href{https://ui.adsabs.harvard.edu/abs/1994ShWav...4...25T}{4, 25}

\bibitem[{{Truelove} {et~al.}(1997){Truelove}, {Klein}, {McKee}, {Holliman},
  {Howell}, \& {Greenough}}]{Truelove1997}
{Truelove}, J.~K., {Klein}, R.~I., {McKee}, C.~F., {et~al.} 1997,
  {\href{https://dx.doi.org/10.1086/310975}{\color{magenta}\apjl}},
  \href{https://ui.adsabs.harvard.edu/abs/1997ApJ...489L.179T}{489, L179}

\bibitem[{{Turk} {et~al.}(2011){Turk}, {Smith}, {Oishi}, {Skory}, {Skillman},
  {Abel}, \& {Norman}}]{Turk2011}
{Turk}, M.~J., {Smith}, B.~D., {Oishi}, J.~S., {et~al.} 2011,
  {\href{https://dx.doi.org/10.1088/0067-0049/192/1/9}{\color{magenta}\apjs}},
  \href{https://ui.adsabs.harvard.edu/abs/2011ApJS..192....9T}{192, 9}

\bibitem[{{Turner} {et~al.}(2017){Turner}, {Consiglio}, {Beck}, {Goss}, {Ho},
  {Meier}, {Silich}, \& {Zhao}}]{Turner2017}
{Turner}, J.~L., {Consiglio}, S.~M., {Beck}, S.~C., {et~al.} 2017,
  {\href{https://dx.doi.org/10.3847/1538-4357/aa8669}{\color{magenta}\apj}},
  \href{https://ui.adsabs.harvard.edu/abs/2017ApJ...846...73T}{846, 73}

\bibitem[{{van Leer}(1979)}]{vanLeer1979}
{van Leer}, B. 1979,
  {\href{https://dx.doi.org/10.1016/0021-9991(79)90145-1}{\color{magenta}J.
  Comput. Phys.}},
  \href{https://ui.adsabs.harvard.edu/abs/1979JCoPh..32..101V}{32, 101}

\bibitem[{{Vasiliev}(2011)}]{Vasiliev2011}
{Vasiliev}, E.~O. 2011,
  {\href{https://dx.doi.org/10.1111/j.1365-2966.2011.18623.x}{\color{magenta}\mnras}},
  \href{https://ui.adsabs.harvard.edu/abs/2011MNRAS.414.3145V}{414, 3145}

\bibitem[{{Veilleux} {et~al.}(2020){Veilleux}, {Maiolino}, {Bolatto}, \&
  {Aalto}}]{Veilleux2020}
{Veilleux}, S., {Maiolino}, R., {Bolatto}, A.~D., \& {Aalto}, S. 2020,
  {\href{https://dx.doi.org/10.1007/s00159-019-0121-9}{\color{magenta}\aapr}},
  \href{https://ui.adsabs.harvard.edu/abs/2020A&ARv..28....2V}{28, 2}

\bibitem[{{Verner} {et~al.}(1996){Verner}, {Ferland}, {Korista}, \&
  {Yakovlev}}]{Verner1996}
{Verner}, D.~A., {Ferland}, G.~J., {Korista}, K.~T., \& {Yakovlev}, D.~G. 1996,
  {\href{https://dx.doi.org/10.1086/177435}{\color{magenta}\apj}},
  \href{https://ui.adsabs.harvard.edu/abs/1996ApJ...465..487V}{465, 487}

\bibitem[{{Verner} \& {Yakovlev}(1995)}]{Verner1995}
{Verner}, D.~A. \& {Yakovlev}, D.~G. 1995, \aaps,
  \href{https://ui.adsabs.harvard.edu/abs/1995A&AS..109..125V}{109, 125}

\bibitem[{{Vink} {et~al.}(2001){Vink}, {de Koter}, \& {Lamers}}]{Vink2001}
{Vink}, J.~S., {de Koter}, A., \& {Lamers}, H.~J.~G.~L.~M. 2001,
  {\href{https://dx.doi.org/10.1051/0004-6361:20010127}{\color{magenta}\aap}},
  \href{https://ui.adsabs.harvard.edu/abs/2001A&A...369..574V}{369, 574}

\bibitem[{{Virtanen} {et~al.}(2020){Virtanen}, {Gommers}, {Oliphant},
  {Haberland}, {Reddy}, {Cournapeau}, {Burovski}, {Peterson}, {Weckesser},
  {Bright}, {van der Walt}, {Brett}, {Wilson}, {Millman}, {Mayorov}, {Nelson},
  {Jones}, {Kern}, {Larson}, {Carey}, {Polat}, {Feng}, {Moore}, {VanderPlas},
  {Laxalde}, {Perktold}, {Cimrman}, {Henriksen}, {Quintero}, {Harris},
  {Archibald}, {Ribeiro}, {Pedregosa}, {van Mulbregt}, \& {SciPy 1. 0
  Contributors}}]{Virtanen2020}
{Virtanen}, P., {Gommers}, R., {Oliphant}, T.~E., {et~al.} 2020,
  {\href{https://dx.doi.org/10.1038/s41592-019-0686-2}{\color{magenta}Nature
  Methods}}, \href{https://ui.adsabs.harvard.edu/abs/2020NatMe..17..261V}{17,
  261}

\bibitem[{{Voronov}(1997)}]{Voronov1997}
{Voronov}, G.~S. 1997,
  {\href{https://dx.doi.org/10.1006/adnd.1997.0732}{\color{magenta}Atom. Data
  Nucl. Data Tabl.}},
  \href{https://ui.adsabs.harvard.edu/abs/1997ADNDT..65....1V}{65, 1}

\bibitem[{{Walsh} \& {Roy}(1989)}]{Walsh1989}
{Walsh}, J.~R. \& {Roy}, J.-R. 1989,
  {\href{https://dx.doi.org/10.1093/mnras/239.2.297}{\color{magenta}\mnras}},
  \href{https://ui.adsabs.harvard.edu/abs/1989MNRAS.239..297W}{239, 297}

\bibitem[{{Weaver} {et~al.}(1977){Weaver}, {McCray}, {Castor}, {Shapiro}, \&
  {Moore}}]{Weaver1977}
{Weaver}, R., {McCray}, R., {Castor}, J., {Shapiro}, P., \& {Moore}, R. 1977,
  {\href{https://dx.doi.org/10.1086/155692}{\color{magenta}\apj}},
  \href{https://ui.adsabs.harvard.edu/abs/1977ApJ...218..377W}{218, 377}

\bibitem[{{Wei{\ss}} {et~al.}(2005){Wei{\ss}}, {Walter}, \&
  {Scoville}}]{Weis2005}
{Wei{\ss}}, A., {Walter}, F., \& {Scoville}, N.~Z. 2005,
  {\href{https://dx.doi.org/10.1051/0004-6361:20052667}{\color{magenta}\aap}},
  \href{https://ui.adsabs.harvard.edu/abs/2005A&A...438..533W}{438, 533}

\bibitem[{{Westmoquette} {et~al.}(2014){Westmoquette}, {Bastian}, {Smith},
  {Seth}, {Gallagher}, {O'Connell}, {Ryon}, {Silich}, {Mayya},
  {Mu{\~n}oz-Tu{\~n}{\'o}n}, \& {Rosa Gonz{\'a}lez}}]{Westmoquette2014}
{Westmoquette}, M.~S., {Bastian}, N., {Smith}, L.~J., {et~al.} 2014,
  {\href{https://dx.doi.org/10.1088/0004-637X/789/2/94}{\color{magenta}\apj}},
  \href{https://ui.adsabs.harvard.edu/abs/2014ApJ...789...94W}{789, 94}

\bibitem[{{W{\"u}nsch} {et~al.}(2011){W{\"u}nsch}, {Silich}, {Palou{\v{s}}},
  {Tenorio-Tagle}, \& {Mu{\~n}oz-Tu{\~n}{\'o}n}}]{Wuensch2011}
{W{\"u}nsch}, R., {Silich}, S., {Palou{\v{s}}}, J., {Tenorio-Tagle}, G., \&
  {Mu{\~n}oz-Tu{\~n}{\'o}n}, C. 2011,
  {\href{https://dx.doi.org/10.1088/0004-637X/740/2/75}{\color{magenta}\apj}},
  \href{https://ui.adsabs.harvard.edu/abs/2011ApJ...740...75W}{740, 75}

\bibitem[{{W{\"u}nsch} {et~al.}(2021){W{\"u}nsch}, {Walch}, {Dinnbier},
  {Seifried}, {Haid}, {Klepitko}, {Whitworth}, \& {Palou{\v{s}}}}]{Wuensch2021}
{W{\"u}nsch}, R., {Walch}, S., {Dinnbier}, F., {et~al.} 2021,
  {\href{https://dx.doi.org/10.1093/mnras/stab1482}{\color{magenta}\mnras}},
  \href{https://ui.adsabs.harvard.edu/abs/2021MNRAS.505.3730W}{505, 3730}

\bibitem[{{W{\"u}nsch} {et~al.}(2018){W{\"u}nsch}, {Walch}, {Dinnbier}, \&
  {Whitworth}}]{Wuensch2018}
{W{\"u}nsch}, R., {Walch}, S., {Dinnbier}, F., \& {Whitworth}, A. 2018,
  {\href{https://dx.doi.org/10.1093/mnras/sty015}{\color{magenta}\mnras}},
  \href{https://ui.adsabs.harvard.edu/abs/2018MNRAS.475.3393W}{475, 3393}

\end{thebibliography}




\end{document}